\documentclass[10pt,journal,compsoc]{IEEEtran}
\ifCLASSOPTIONcompsoc
  \usepackage[nocompress]{cite}
\else
  \usepackage{cite}
\fi
\ifCLASSINFOpdf
\else
\fi
\usepackage{url}
\usepackage[pdftex]{graphicx}
\usepackage{epstopdf}
\usepackage{booktabs} % For formal tables
\usepackage{tabularx,multirow,booktabs,blindtext}
\usepackage[para]{threeparttable}
\usepackage{colortbl}
\usepackage[table]{xcolor}

\newcommand{\figref}[1]{Fig.~\ref{#1}}

\begin{document}

%
% paper title
% Titles are generally capitalized except for words such as a, an, and, as,
% at, but, by, for, in, nor, of, on, or, the, to and up, which are usually
% not capitalized unless they are the first or last word of the title.
% Linebreaks \\ can be used within to get better formatting as desired.
% Do not put math or special symbols in the title.
\title{A Transprecision Floating-Point Cluster for Efficient Near-Sensor Data Analytics}
%
%
% author names and IEEE memberships
% note positions of commas and nonbreaking spaces ( ~ ) LaTeX will not break
% a structure at a ~ so this keeps an author's name from being broken across
% two lines.
% use \thanks{} to gain access to the first footnote area
% a separate \thanks must be used for each paragraph as LaTeX2e's \thanks
% was not built to handle multiple paragraphs
%
%
%\IEEEcompsocitemizethanks is a special \thanks that produces the bulleted
% lists the Computer Society journals use for "first footnote" author
% affiliations. Use \IEEEcompsocthanksitem which works much like \item
% for each affiliation group. When not in compsoc mode,
% \IEEEcompsocitemizethanks becomes like \thanks and
% \IEEEcompsocthanksitem becomes a line break with idention. This
% facilitates dual compilation, although admittedly the differences in the
% desired content of \author between the different types of papers makes a
% one-size-fits-all approach a daunting prospect. For instance, compsoc
% journal papers have the author affiliations above the "Manuscript
% received ..."  text while in non-compsoc journals this is reversed. Sigh.

\author{Fabio~Montagna,
        Stefan~Mach,
        Simone~Benatti,
        Angelo~Garofalo,
        Gianmarco~Ottavi,
        Luca~Benini,~\IEEEmembership{Fellow,~IEEE,}
        Davide~Rossi,~\IEEEmembership{Member,~IEEE,}
        and Giuseppe~Tagliavini,~\IEEEmembership{Member,~IEEE}% <-this % stops a space
\IEEEcompsocitemizethanks{\IEEEcompsocthanksitem F.~Montagna and G.~Tagliavini are with the Dept. of Computer Science and Engineering (DISI), University of Bologna, Italy.\protect\\
E-mail: \{fabio.montagna, giuseppe.tagliavini\}@unibo.it% <-this % stops an unwanted space
\IEEEcompsocthanksitem S.~Mach and L.~Benini are with the Dept. of Information Technology and Electrical Engineering (D-ITET), ETH Z\"urich, Switzerland.\protect\\
E-mail: \{smach, benini\}@iis.ee.ethz.ch% <-this % stops an unwanted space
\IEEEcompsocthanksitem A.~Garofalo, G.~Ottavi, S.~Benatti, L.~Benini and D.~Rossi are with the Dept. of Electrical, Electronic, and Information Engineering (DEI), University of Bologna, Italy.\protect\\
E-mail: \{angelo.garofalo, gianmarco.ottavi2, simone.benatti, davide.rossi, luca.benini\}@unibo.it}
\thanks{This work has been partially supported by the European Union’s Horizon 2020 research and innovation programme under grant agreement numbers 732631 (OPRECOMP) and 857191 (IOTWINS).}}

\IEEEtitleabstractindextext{%
\begin{abstract}
Recent applications in the domain of near-sensor computing require the adoption of floating-point arithmetic to reconcile high precision results with a wide dynamic range.
In this paper, we propose a multi-core computing cluster that leverages the fined-grained tunable principles of transprecision computing to provide support to near-sensor applications at a minimum power budget.
Our design -- based on the open-source RISC-V architecture -- combines parallelization and sub-word vectorization with near-threshold operation, leading to a highly scalable and versatile system.
We perform an exhaustive exploration of the design space of the transprecision cluster on a cycle-accurate FPGA emulator, with the aim to identify the most efficient configurations in terms of performance, energy efficiency, and area efficiency.
We also provide a full-fledged software stack support, including a parallel runtime and a compilation toolchain, to enable the development of end-to-end applications.
We perform an experimental assessment of our design on a set of benchmarks representative of the near-sensor processing domain, complementing the timing results with a post place-\&-route analysis of the power consumption.
Finally, a comparison with the state-of-the-art shows that our solution outperforms the competitors in energy efficiency, reaching a peak of 97 Gflop/s/W on single-precision scalars and 162 Gflop/s/W on half-precision vectors.
\end{abstract}

% Note that keywords are not normally used for peerreview papers.
}

% make the title area
\maketitle

% To allow for easy dual compilation without having to reenter the
% abstract/keywords data, the \IEEEtitleabstractindextext text will
% not be used in maketitle, but will appear (i.e., to be "transported")
% here as \IEEEdisplaynontitleabstractindextext when the compsoc
% or transmag modes are not selected <OR> if conference mode is selected
% - because all conference papers position the abstract like regular
% papers do.
\IEEEdisplaynontitleabstractindextext
% \IEEEdisplaynontitleabstractindextext has no effect when using
% compsoc or transmag under a non-conference mode.

 \IEEEraisesectionheading{\section{Introduction}}
\label{sec:intro}
\IEEEPARstart{T}{he} pervasive adoption of edge computing is increasing the computational demand for algorithms targeted on embedded devices.
Besides the aggressive optimization strategies adopted on the algorithmic side \cite{tagliavini2018transprecision}, there is a great effort to find the best trade-off between architectural features and computational capabilities \cite{pullini2019mr}. Indeed, deploying artificial intelligence algorithms or digital signal processing (DSP) on near-sensor devices poses several challenges to resource-constrained low-power embedded systems.
%
%For instance, in the domains of Internet of Things (IoT) and Machine Learning (ML), the main challenge is the enhancement of high-level applications for the deployment on embedded devices, which have several limitations in maximum operative frequency and battery duration.
%
%Fixed-point arithmetic has played an important role in embedded systems optimization \textbf{(citazioni)}, by virtue of its efficiency simplicity of the architectures and the energy benefits derived from the use of integer arithmetic units over floating-point (FP) units.

Fixed-point arithmetic is a well-established paradigm in embedded systems optimization since it allows a simplified numerical representation for real numbers at high energy efficiency \cite{barrois2017customizing}.
Nevertheless, many applications require high precision results characterized by a wide dynamic range (e.g., the accumulation stage of support vectors, or feed-forward inference for deep neural networks).
%Nevertheless, some deploying computationally intensive ML tasks in near-sensor processing applications, such as bio-potential analysis, drone control, or robotics often requires high dynamic ranges and strict numerical precision.
%
%The deployment of computational intensive ML or linear algebra algorithms in near-sensor processing applications (e.g., bio-potential analysis, drone control, or robotics) often requires high dynamic ranges and strict numerical precision.
%
In these cases, fixed-point implementations may suffer from numerical instability, requiring an in-depth analysis to make the result reliable and additional code sections to normalize and adjust the dynamic range avoiding saturation (e.g., the fixed-point implementation of linear time-invariant digital filters described in \cite{volkova20}).
As a result, fixed-point arithmetic is not necessarily the most energy-efficient solution since the code requires real-time adaptations of the dynamic range that affect performance significantly and increase the time-to-market \cite{barrois2017customizing}.
To cope with these issues, the adoption of single-precision floating-point (FP) arithmetic is a well-established paradigm for embedded low-power systems, such as ARM Cortex M4, a microcontroller (MCU) architecture that is the \textit{de facto} standard for FP-capable low-power edge nodes. Combining FP and fixed-point arithmetic, depending on the computational requirements, is the typical approach for optimizing Cortex-M4 applications.
%CMSIS DSP libraries provide software functions to convert numerical representations (i.e., fixed-to-float and \textit{viceversa}), resulting in better performance in terms of execution time for complex algorithms.
The main shortcomings of this approach are related to the manual analysis for the format selection (float vs. fixed), the tuning required for adjusting the fixed-point dynamic range, and the software overhead to make the format conversions. Furthermore, usage of mixed (i.e., floating-fixed point) representation introduces several architectural bottlenecks in managing the pipelines and the register file, such as flushing or stalls that reduce the computational efficiency of these approaches. Finally, at least in commercial architectures, the floating-point unit (FPU) cannot be turned off while the core is executing fixed-point operations, resulting in a further reduction of energy efficiency.
 
%The adoption of the standard double- or single-precision FP formats provides to fulfill the typical accuracy constraints of most near-sensor applications.
%
 %However, these formats require a non-negligible computational effort on deeply-embedded processors, where the minimization of power consumption is a fundamental aspect of the system design.
%
%
%\textit{Transprecision computing} is gaining ground in recent years, as the variety of the algorithms implemented in embedded systems is becoming more and more computationally intensive, also requiring high numerical precision.
%
%For instance, in the era of the Internet of Things (IoT) and Machine Learning (ML), the main challenge is the enhancement of high-level applications, for the deployment on embedded devices, which have several limitations in maximum operative frequency and battery duration.
%
In this scenario, \textit{transprecision computing} \cite{malossi2018transprecision} is emerging as a successful paradigm for embedded computing systems. This paradigm is an evolution of approximate computing, and it aims at tuning approximation at a fine grain during the computation progress through hardware and software control mechanisms.
%
%On the hardware side, this approach requires computing architectures that operate with a smooth and wide range of precision vs. cost trade-off curves.
In the context of FP computations, this approach requires the availability of hardware units providing efficient support for multiple FP formats.
%
%Most FPUs represent real numbers following the IEEE 754 data format. This specification also includes a 16-bits format (half-precision) that allows implementing SIMD-like operations to further improve performance, reducing the memory accesses and parallelizing the computation.
%
The IEEE 754-2008 standard \cite{zuras2008ieee} describes five FP formats, and two are suitable for our target: the 16-bits half-precision format (\emph{float16}) and the 32-bits single-precision format (\emph{float}).
Moreover, \emph{bfloat16} is an alternative 16-bit format dedicating 8 bits to the exponent field, in contrast with the 5 bits used in the IEEE half-precision format; this trade-off allows to handle the same dynamic range of the \emph{float} format losing some precision. % a numerical dynamic equivalent to the \emph{float} format losing some precision.
Relevant applications in the field of near-sensor computing \cite{tagliavini2018transprecision} as well as state-of-the-art (SoA) machine learning algorithms widely rely on \emph{float16} and \emph{bfloat16} formats since many ML models, such as Temporal Convolutional Networks (TCN) \cite{8930945} or Convolutional Neural Networks (CNN) \cite{wu2016quantized}, tolerate lower precision arithmetic without losing their accuracy\cite{burgess2019bfloat16}.
Adopting the smaller format that satisfies the application accuracy requirements paves the way for substantial improvements in performance and energy consumption. Still, programmers need full software support in the compilation toolchain and also a consolidated methodology for tuning the precision of FP variables \cite{tagliavini2018flexfloat}.
Table \ref{tab:fp_formats} provides an overview of these formats.
\begin{table}[t]
\caption{Floating-point formats used in low-power embedded systems.}
\label{tab:fp_formats}
\resizebox{\columnwidth}{!}{%
\setlength{\tabcolsep}{2pt}
\begin{threeparttable}
\begin{tabular}{|l|c|c|c|c|}
\hline
\textbf{Format\tnote{1}}   & \textbf{Exponent} & \textbf{Mantissa}  & \textbf{Range} & \textbf{Accuracy\tnote{2} } \\ 
\hline
\emph{\textbf{float}}    & 8  & 23 & $1.2\times10^{-38}$ -- $3.4\times10^{38}$ & 7.2 \\
\hline
\emph{\textbf{bfloat16}} & 8  & 7  & $1.2\times10^{-38}$ -- $3.4\times10^{38}$ & 2.4 \\
\hline
\emph{\textbf{float16}}  & 5  & 11 & $5.9\times10^{-88}$ -- $6.5\times10^{4}$ & 3.6\\
\hline
\end{tabular}
\end{threeparttable}}
\begin{tablenotes}
\item[1] Number of bits.
\item[2] Decimal digits.
\end{tablenotes}
\end{table}
%
%
%(\textbf{!"})For some applications, the precision can be further reduced, defining other data types. (\textbf{"!})
%
% In \cite{tagliavini2018transprecision} the authors present an FP type system with complete hardware support for the standard formats, including also non standard 8-bit and 16-bit formats.
%
% Moreover, they present a methodology to integrate the FP library with an external tool for precision tuning, demonstrating the benefits of using these format and the tool in several DPS benchmarks.
% %
% The \textit{binary16alt} allows to handle a numerical dynamic equal to the one provided by the \textit{binary32}, because the number of bits dedicated to the exponent is the same, but loosing precision as a lower number of bits is dedicated to the mantissa.
%
%It has been demonstrated that many models for ML tolerate lower precision arithmetic, without loosing (or even increasing) the accuracy.
%(\textbf{TensorFlow doc: https://cloud.google.com/tpu/docs/bfloat16}).
%
%The use of this format paves the way to substantial improvements in performance, using vectorization to speed-up the execution, and memory usage, allowing to store bigger models in the same amount of memory.
%
%
%\textbf{CENNO ALLE gpu USATO PER IL TRAINING (bFLOAT) ADN GOOGLE}

Lowering the bitwidth of FP variables paves the way for crucial optimizations. The most relevant one is the possibility of performing instructions on multiple sub-word elements simultaneously using a single-instruction-multiple-data (SIMD) approach on 16-bits vectors. These data types are known as \emph{packed-SIMD vectors}.
SIMD operations act on multiple data elements of the same size and type simultaneously, offering a theoretical $2\times$ speed-up for 16-bits data. 
Moreover, vectorization of 16-bits types enables an equivalent reduction of the memory footprint, allowing to store bigger models in the same amount of memory.
This approach also enables more effective data movements, as multiple elements can be transferred concurrently.

%\textbf{Parallel near-threshold computing – cluster efficienti TC bla bla.}
%
An architectural design that aims to achieve the goals discussed above must exploit additional features of the ultra-low-power (ULP) computing domain.
A tightly-coupled cluster composed of several processing elements (PEs) enables to improve the computational capabilities of the system using parallel programming techniques, without increasing the operating frequency.
Specialized hardware extensions allow programmers to accelerate key parallel patterns and exploit the advantages of packed-SIMD operations.
Combining these features with near-threshold computing on a fully programmable multi-core architecture leads to a highly scalable and versatile system suitable for a wide range of applications.
%
%\textbf{-	Quando abbiamo app FP servono istruzioni per portare dati all’interno delle unità aritmetiche, Von Neumann bottleneck  e molto spesso è poco efficiente dai punti di vista dell’area avere una unità per core.}
%
The number of FPUs and the sharing factor among the cores of the cluster require a careful evaluation since these aspects directly impact area and energy efficiency.
For instance, having a dedicated FPU for the cores can be detrimental if the data demand from the PEs can not be satisfied by the memory throughput: this effect is known as the Von Neumann bottleneck.
Thus, reducing the number of FPUs and adopting a sharing policy among the cores can be beneficial to improve the system's efficiency.
Another aspect to consider is the pipelining of the FPU unit, which allows designers to increase the maximum operating frequency to the cost of a deterioration of the performance.
In the depicted scenario, finding the best trade-off requires an accurate exploration of the design space that includes the definition of adequate metrics and an experimental assessment on kernels from end-to-end applications.

In this paper, we propose the design of a \emph{transprecision computing cluster} tailored for applications in the domain of near-sensors computing.
Our work includes the architectural design of the transprecision cluster and the full specification of FPU and interconnects.
We also provide full-fledged software support for the transprecision cluster, including a runtime environment for parallel programming and an extended compilation toolchain.
We performed a comprehensive exploration of the design space considering different architectural configurations of the transprecision cluster: the number of cores, the number of FPUs and the related sharing factor, the number of pipeline stages in the FPUs. We have performed this exploration on an accurate hardware emulator implemented on an FPGA board. We have performed experiments on a set of benchmarks, including kernels commonly used in end-to-end applications identifying the most efficient solutions in terms of performance, power, and area; for this evaluation, we have considered the experimental results and power/area figures derived from post place-\&-route (P\&R) models in 22nm FDX technology. Based on this exploration, we derive guidelines to instantiate an optimal proper cluster configuration depending on the target application domain and expected performance. Finally, we have compared the most efficient configurations deriving from the design space exploration with SoA solutions, considering a broader scenario that includes high-performance and embedded computing domains.

Our experimental results show that the configuration with 16 cores and private FPUs configured with one pipeline stage provide the best performance (5.92 Gflop/s), the one with 16 cores and private FPUs configured with zero pipeline stages is the most energy-efficient (167 Gflop/s/W), and finally the configuration with 8 cores and 4 shared FPUs configured with one pipeline stage is the most area-efficient (3.5 Gflop/s/mm$^2$).
Finally, the energy efficiency of the transprecision cluster outperforms all the other solutions that provide FP support in the area of embedded computing.

The rest of the paper is organized as follows: Section \ref{sec:related} presents the related work.
The proposed architecture of the transprecision cluster is presented in Section \ref{sec:cluster}.
Section \ref{sec:software} describes the programming model and the compilation toolchain.
Sections \ref{sec:experiments} and \ref{sec:soa} present the experimental results and a comparison with previous works, respectively.
Finally, Section \ref{sec:conclusion} concludes the whole paper.
 \section{Related Work}
\label{sec:related}

\subsection{Alternative formats}
The IEEE754-2008 standard includes the definition of decimal formats for contexts where the accumulation of rounding errors in binary-radix numbers can lead to unacceptable accuracy losses.
This effect is due to the fact that a set of finite radix-10 numbers becomes periodic when represented in radix-2 notation.
Wahba et al. \cite{wahba2016area} present a solution reducing by 6\% percent the latency of an FP decimal unit compared to SoA solutions, and saving 23\% of the total area compared to solutions that include two FP units (for binary and decimal support, respectively).
Decimal FPUs are characterized by a longer critical path and a larger area than binary units since representing a decimal digit requires four bits. For these reasons, they are not a suitable alternative for integration in our transprecision cluster.

In recent years, researchers have started to explore custom formats that are alternative to the IEEE standard ones and its closer derivatives.
% LNU
Gautschi et al. \cite{gautschi2017shared} propose an FPU based on the logarithmic number system (LNU), which is up to $4.1\times$ more energy efficient than standard FP units in non-linear processing kernels.
%Even though the presented results are quite promising, not all FP operations can be easily implemented, and the practical benefits of an LNU are domain-specific.
% UNUM
Universal numbers (\emph{UNUMs}) \cite{gustafson2017end} adopt a variable-width representation based on interval arithmetic to guarantee that the result contains the exact solution \cite{glaser2018unum}.
The variable width provided by UNUM enables to scale up precision in scientific computing applications \cite{bocco2019smurf}, but hardware implementations not suitable for the area and energy constraints of the embedded computing domain. 
A recent version of the UNUM specification, known as \emph{UNUM type III} or \emph{posit} \cite{gustafson2017beating}, proposes a solution to the hardware overhead issue.
\cite{glaser2018unum} introduces a posit arithmetic unit supporting additions and subtractions, coupled with dedicated compression units to reduce the memory footprint.
%The synthesis in a 65 nm CMOS process achieves a maximum clock frequency of 413 MHz at 1.2 V with an average measured power of 210 uW/MHz and an area footprint of 0.258 mm$^2$.
%Even though this solution can be considered suitable for the embedded domain, the authors conclude that this format provides limited benefits in reducing the memory footprint compared to the IEEE 754 formats (-7\%). At the same time, the datapath complexity is significantly higher since it includes the logic to compute upper and lower bounds for interval arithmetic and the support for on-the-fly data compression.
%
Considering the overhead of current hardware implementations and their limited benefits (in \cite{glaser2018unum}, authors estimate a reduction of memory footprint around 7\%), counterpoised with a significant effort in code refactoring, in this work we have not considered these hardware components as viable candidates for the transprecision cluster design.

\subsection{Transprecision computing building blocks}
The choice of an energy-efficient and transprecision-enabled FPU design is a key enabler for this work.
In literature, there are several designs of FPUs that enable transprecision operations. 
For instance, Kaul et al. \cite{kaul2012varprec} describe a variable-precision fused multiply-and-add (FMA) unit with vector support (1, 2, or 4 ways). Their design considers 8 bits for the exponents and 24 bits for the mantissa. Moreover, a 5-bits certainty field tracks the number of accurate mantissa bits: Computations that do not fulfill the accuracy constraints provided by the application are recomputed with increased precision. 
The energy consumption for a 32 nm CMOS implementation is 19.4 pJ/FLOP, even though the overhead due to precision tracking and fixed-size exponents increases the total energy consumption at the application level.
Moreover, if maximum precision is required, applications become very inefficient, due to the need for repeated operations performed at a lower precision.

%\cite{nannarelli2018tunable}
Nannarelli \cite{nannarelli2019tunable} describes the design of an FPU based on the Tunable Floating-Point (TFP) format, which supports a variable number of bits for mantissa (from 4 to 24) and exponent (from 5 to 8).
%Considering an FPU containing both units synthesized in a 45 nm CMOS library, the estimated power consumption is around 20.18 mW (at 1 GHz), and the area around 15805 um$^2$.
However, this solution does not support vectorization, which is a key enabler for energy efficiency.
Jaiswal et al. \cite{Jaiswal2015} present a pipelined design of two FP adders that support multiple precision configurations.
%The first architecture can operate either for a quadruple-precision scalar or a double-precision 2-way vector, while the second one supports a range from quadruple-precision scalar computations to single-precision 4-way vectorial computations. 
%A pipelined implementation of both adder architectures based on UMC 90 nm technology is presented. 
The results are promising in terms of area and energy efficiency, but this solution does not support additional FP operations.
Hardfloat \cite{asanovic2016rocket} is an open-source library (written in Chisel) that contains parameterized blocks for FMA operations, conversions between integer and FP numbers, and conversions among different FP formats.
At the current stage of development, this library offers individual function blocks instead of a fully-featured FPU, missing unit-level optimizations.
Zhang et al. \cite{zhang19eff} present a multiple-precision FP FMA with vector support
%to execute one quadruple-precision operation, two parallel double-precision operations, four parallel single-precision operations, or eight parallel half-precision operations.
Their work aims at minimizing the area overhead, but the hardware sharing inside the datapath constrains all formats to use the same latency. Moreover, the FPU does not provide any support for scalars in smaller formats.

FPnew \cite{mach2020fpnew} is an open-source transprecision floating-point unit (TP-FPU) capable of supporting a wide range of standard (double, float, and float16) and custom (bfloat16 and 8-bit minifloat) FP formats.
%This FPU supports both scalar and packet-SIMD operations, achieving high energy efficiency (178 Gflop/sW on doubles and 2.95 Tflop/sW on 8-bit mini-floats) and performance (between 3.2 Gflop/s and 25.3 Gflop/s).
FPnew supports both scalar and packet-SIMD operations, and the experimental results shown in \cite{mach2020fpnew} assess that this design outperforms all its competitors in terms of area and energy efficiency.
We have integrated this FPU in our architecture, Section~\ref{sec:cluster_fpu} describes its design and integration aspects in further detail.
%Bruguera \cite{bruguera2019low} describes low-latency FP divider and square root units based on Digit-recurrence algorithms.
%These units require 11, 6, and 4 cycles for double, single, and half-precision division, respectively, and 15, 8, and 5 cycles for square root; the presence of subnormal operands or result requires two additional cycles.
FPNew includes a DIVSQRT module to compute divisions and square roots using an iterative non-restoring divider, similar to the design presented in \cite{bruguera2019low}.
%However, we use a different sharing factor for this block since these operations are much less frequent than FMA in typical use cases.

\subsection{Vector units}
Variable-length vector units have been initially introduced on the CRAY-1 architecture \cite{russell1978cray}, and today they are a well-established solution in the high-end computer systems.
The ARM Scalable Vector Extension (SVE) \cite{stephens2017arm} is a vector extension introduces as a SIMD instruction set for the AArch64 architecture.
The SVE specification allows system designers to choose a vector register length between 128 and 2,048 bits to satisfy different constraints.
The programming model is vector-length agnostic; there is no need to recompile the source code or use compiler intrinsics to change the vector length.
The A64FX chip by Fujitsu is realized in TSMC 7nm technology and implements the SVE extension, including 48 cores with support for 512-bit wide vectors and reaching peak performance of 2.7 Tflop/s \cite{yoshida2018fujitsu}.
This chip has been used in Fugaku, which entered the TOP500 list in June 2020 as the fastest supercomputer in the world. 

The current working draft for the RISC-V ‘V’ vector extension\footnote{https://github.com/riscv/riscv-v-spec} defines a variable-length register file with vector operation semantics.
This extension provides support for FP16, FP32, FP64, and FP128 types, and also includes widening FMA operations to support mixed-precision computations (e.g., multiplying two FP16 registers and adding the result to an FP32 register).
%
%ARA \cite{cavalcante2019ara} is a chip implemented in GLOBALFOUNDRIES 22FDX FD-SOI technology and including an Ariane core and a 64-bit vector unit based on the version 0.5 draft of RISC-V vector extension. An instance with sixteen lanes achieves up to about 41 GFLOP/s/W (in double precision) at a nominal frequency of 1.04 GHz with an area of 10735 kGE (2.14 mm$^2$).
%
%Hwacha \cite{dabbelt2016vector} is a vector core based on the vector-fetch [12] paradigm to an array of single-issue, in-order RISC-V Rocket cores \cite{lee201445nm}.
%A Hwacha instance in a 28 nm technology with four lanes has an area of 1.226 mm$^2$ and a power consumption 481 mW at 850 MHz.
ARA \cite{cavalcante2019ara} and Hwacha \cite{lee201445nm} are two embodiments of this provisional standard.
ARA includes an RV64 core and a 64-bit vector unit based on the version 0.5 draft of the RISC-V vector extension.
Hwacha is based on the vector-fetch paradigm and is composed of an array of single-issue, in-order RISC-V Rocket cores \cite{lee201445nm}.
In general, the area and power consumption of these solutions are too high for low-power, MCU-class processing systems. This observation is the main reason why vector semantics in ULP embedded systems are typically supported by providing packed-SIMD instructions.

\subsection{Software-based transprecision approaches}
Besides approaches involving custom HW design to enable mixed-precision operations, several researchers have proposed multiple-precision arithmetic libraries that extend the IEEE754 formats to perform FP computations with arbitrary precision. This solution allows application designers to overcome the limitations of fixed-format FP types without dedicated hardware support.
ARPREC \cite{bailey2002arprec} and GNU MPFR \cite{fousse2007mpfr} provide APIs to handle multiple formats characterized by a fixed-size exponent (a machine word) and an arbitrary size mantissa (multiples of a machine word).
Arb \cite{johansson2017arb} is a C library for arbitrary-precision interval arithmetic using the midpoint-radius representation that outperforms non-interval solutions such as MPFR in some applications.
These libraries are widely used in contexts requiring high-dynamic range and are characterized by relaxed constraints on computation time and energy consumption (e.g., scientific computing on data center nodes).
To speed up the library execution time, {Lef\`e}vre \cite{lefevre2017correctly} presents a new algorithm to speed up the sum of arbitrary-precision FP number using the MPRF internal representation,
However, the approach based on software emulation is not a viable solution for energy-efficient embedded systems, since both time and energy efficiency are negatively affected by at least an order of magnitude compared with solutions based on dedicated hardware.

Anderson et al. \cite{anderson2017efficient} propose a software approach for the reduced-precision representation of FP data.
They define a set of non-standard floating-point multibyte formats (flytes) that can be converted to the next largest hardware type to perform arithmetic computations.
The exponent is set to the maximum number of bits of the containing type to minimize the conversion overhead.
The adoption of the vector units available on general-purpose processors (e.g., Intel Haswell) or high-end accelerators (e.g., Intel Xeon Phi) allows the software library to coalesce memory accesses and then amortize the conversion overhead.

\subsection{Low-power parallel architectures for FP computing}
Coarse Grain Reconfigurable Architectures (CGRAs) recently emerged as a promising solution for the near-sensor processing domain.
CGRAs are systolic arrays containing a large number of processing elements with a low-latency routing interconnect.
MuTARe \cite{brandalero2020multi} is a CGRA working in the near-threshold voltage regime to achieve low energy.
This solution improves by 29\% the energy efficiency of a heterogeneous platform based on the ARM big.LITTLE platform.
However, MuTARe targets high-end embedded systems, and it does not provide FP support.
Transpire \cite{prasad2020transpire} is a CGRA architecture with FP support.
This architecture has a potential performance improvement of 10.06$\times$ and can be 12.91$\times$ more energy efficient than a RISC-V core extended with packed-SIMD vectorization.
These benefits are due to the fact that the design of CGRAs enables programmers to exploit different combinations of data-level and pipeline-based parallelism.
However, the domain of near-sensor processing includes a wide variety of algorithms presenting complex access patterns that cannot be efficiently implemented on CGRAs.

Mr.Wolf \cite{pullini2019mr} is a multi-core programmable processor implemented in CMOS 40nm technology. The platform includes a tiny (12 Kgates) RISC-V core accelerated by a powerful 8-cores cluster of RI5CY cores \cite{gautschi2017near} sharing two single-precision FPUs.
The limited number of FPUs provided by this architecture represents a severe limitation to the maximum FP intensity that applications may expose.
A primary contribution of our work consists of
finding the best tradeoff between the number of cores and the number of available FPUs, yet considering strict area and power constraints, and exploiting transprecision units to improve performance and execution efficiency.
In Section~\ref{sec:soa}, we include Mr.Wolf in our comparison with SoA platforms.

Helium\footnote{\url{https://www.arm.com/why-arm/technologies/helium}} is an extension of the Armv8.1-M architecture targeting low-power MCU-class computing systems.
The ISA extension includes a set of scalar and vector instructions supporting fixed-point (8-bit, 16-bit, and 32-bit) and FP (float and float16, optionally double) formats.
These instructions are beneficial for a wide range of near-sensor applications, from machine learning to DSP.
The Cortex-M55\footnote{\url{https://www.arm.com/products/silicon-ip-cpu/cortex-m/cortex-m55}} core includes the Helium extension, but chips based on this IP are not yet available on the market to perform a comparison with our solution.

\subsection{High-end embedded systems for FP computing}
The most widely used commercial architectures for compute-intensive FP workloads are GP-GPUs. With the growth of emerging applications such as training of neural networks, they have also started to support reduced precision floating-point formats such as \emph{brain-float} and \emph{binary16}. Indeed, training algorithms for deep neural networks such as backpropagation is naturally robust to errors. These features of modern GPUs have also been exploited in other application domains, such as machine learning \cite{ho2017exploiting} and linear algebra \cite{eliuk2016dmath}, demonstrating significant benefits for performance and efficiency. NVidia Pascal has been the first GPU supporting 16-bit FP formats. NVidia Pascal features SIMD float16 operations, that can be executed using a single paired-operation instruction. Furthermore, the new Volta micro-architecture further extends support to reduced precision types featuring mixed-precision multiply-and-add instructions. 

%From a research point of view, some recent research works, such as the one of Mukunoki et al. \cite{mukunoki2016reducedgpu} explored new custom FP formats for GPUs, demonstrating that performance can be improved significantly exploiting small-bit-width data formats while forcing data to be word-aligned in memory, considerably reducing the number of memory accesses and the related energy cost.
%
Other research works, targeting neural network training and FP intensive workloads, leverage more specialized architectures. Neurostream \cite{azarkhish2017neurostream} is a streaming memory-mapped co-processor targeting inference and training of deep neural networks in near-memory computing systems. This design removes the register-file bottleneck of SIMD architectures accessing the memory exploiting programmable hardware loops and address generators, and enabling execution efficiency close to one MAC per cycle. Neurostream achieves an average performance of 240 Gflop/s within a power-budget of 2.5 W. The architecture has been further improved in \cite{schuiki2018scalable}, with a 2.7$\times$ energy efficiency improvement over GPGPUs at 4.4$\times$ less silicon area, delivering 1.2 TFLOP/s. The latter architecture has been implemented in 22nm FDX in Kosmodrom \cite{zaruba2019floating}. The chip includes two Ariane cores and one NTX accelerator. Kosmodrom achieves an energy efficiency of 260 Gflop/s/W and a 28 Gflop/s performance within a 6.2 mW to 400 mW power envelope.

In \cite{zaruba2020snitch}, the memory-mapped control has been replaced by a tiny general-purpose processor meant to drive double-precision FPUs, improving efficiency and flexibility of previous approaches. This architecture introduces two ISA extensions to reduce the pressure on the core: the stream semantic registers (SSR) and the floating-point repetition instruction (FREP). SSRs allow the core to implicitly encode memory accesses as register reads/writes, removing a significant number of explicit memory instructions. The FREP extension decouples the FP and integer pipeline by sequencing instructions from a micro-loop buffer. The evaluation on an octa-core cluster in 22 nm technology reports a 5$\times$ multi-core speed-up and a 3.5$\times$ gain in energy efficiency.

The architectures discussed in this section target the domain of servers and high-end embedded systems, and presenting further details is beyond the scope of our work. However, the comparison with these solutions provides useful insight and is discussed in Section~\ref{sec:soa}.

 \begin{figure}[t]
\begin{center}
\includegraphics[width=\linewidth]{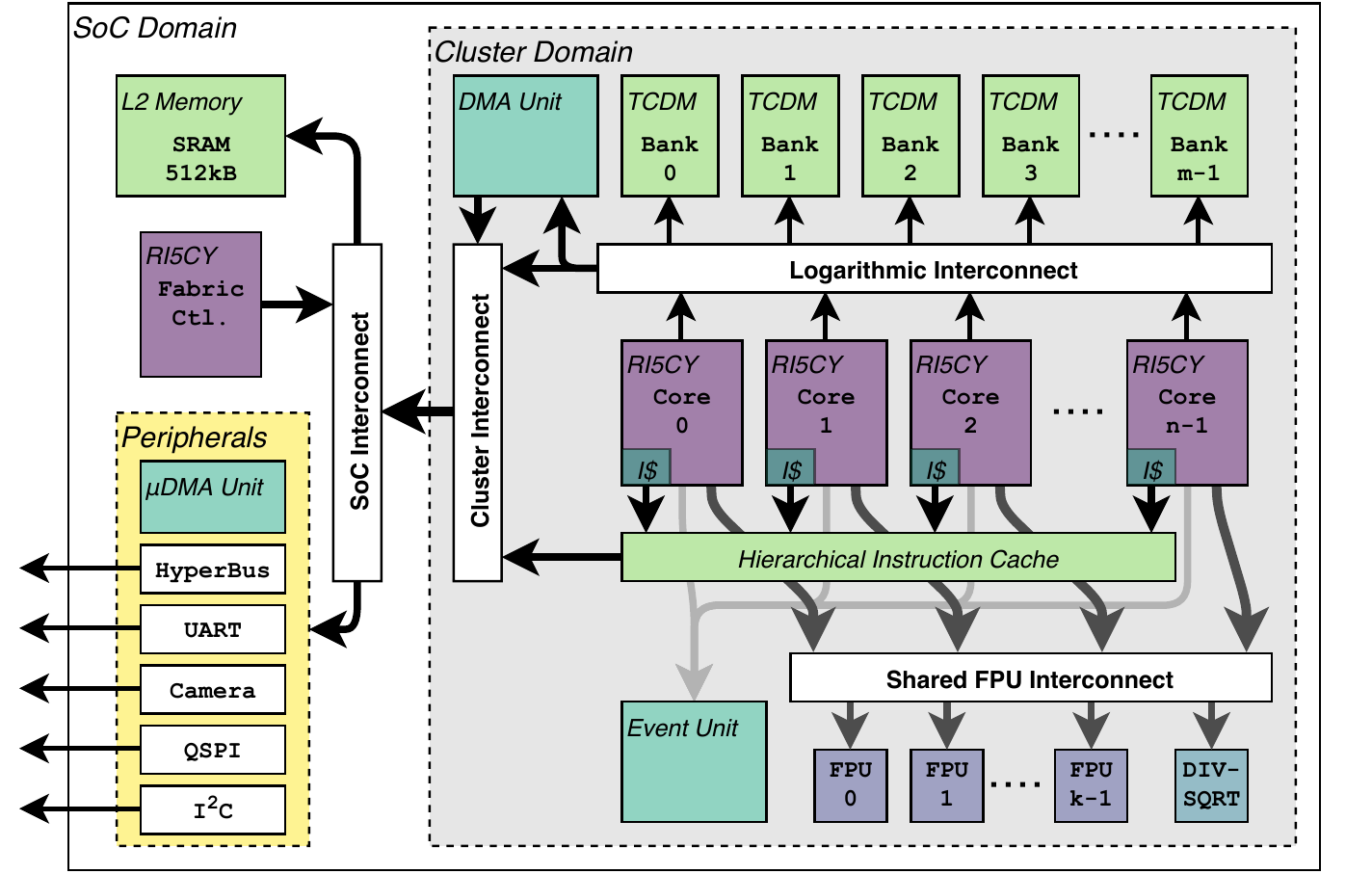}
%\vspace{-6mm}
\caption{Top-level view of the proposed transprecision cluster.}
\vspace{-3mm}
\label{fig:cluster}
\end{center}
\end{figure}
\section{Architecture and Implementation}
\label{sec:cluster}

\subsection{Cluster Architecture}
\label{sec:cluster_archi}
The cluster architecture proposed in this work is a soft IP implementing a tightly-coupled cluster of processors built around a parametric number of RI5CY \cite{gautschi2017near} cores.
Fig.\ref{fig:cluster} shows the top-level design of the transprecision cluster.

RI5CY is a RISC-V based processor implementing a 4-stage in-order single-issue pipeline, supporting the RV32IMC instruction set and dedicated extensions for DSP and machine learning workloads \cite{gautschi2017near}.
The cores fetch instructions from a 2-levels shared instruction cache optimized for performance and energy efficiency when running SIMD workloads typical of near-sensor data analytics applications.
To enable the single-cycle exchange of data among cores, they share a multi-banked Tightly-Coupled Data Memory (TCDM) behaving as a scratchpad memory (i.e., there is no data caching mechanism to avoid coherency and control overheads).
The TCDM enables the cores to share data through a word-level interleaved, single-cycle latency logarithmic interconnect, allowing the execution of data-parallel programming models such as OpenMP. A dedicated hardware block (Event Unit) provides low-overhead support for fine-grained parallelism, accelerating the execution patterns typical of data-parallel programming models (e.g., thread dispatching, barriers, and critical regions) and enabling the adoption of power-saving policies when cores are idle \cite{glaser2019}.

Outside the cluster, at the SoC level, the architecture features one more memory hierarchy level, composed of a 15-cycle latency multi-banked scratchpad memory used to serve the data bus of the cores, the instruction cache refills, and the cluster DMA. We base the explorations performed in this work on a set of cluster configurations with 8 and 16 cores. The L2 memory comprises 512 kB, the TCDM is 64 kB for the 8-core configurations and 128 kB for the 16-core ones. The cluster cores are connected to multiple FPU instances, whose number and interconnect are a central part of our exploration.
Unlike the standard configuration for the RI5CY core, the proposed cluster does not employ core-private FPUs.
Instead, a set of FPUs is shared among all cores in the cluster, using an interconnect which enables various mappings of cores to available FPUs.
The next section provides insights into the FPU subsystem proposed in this work.

\subsection{FPU and interconnect}
\label{sec:cluster_fpu}
\begin{figure}[t]
\centering
\includegraphics[width=0.95\linewidth]{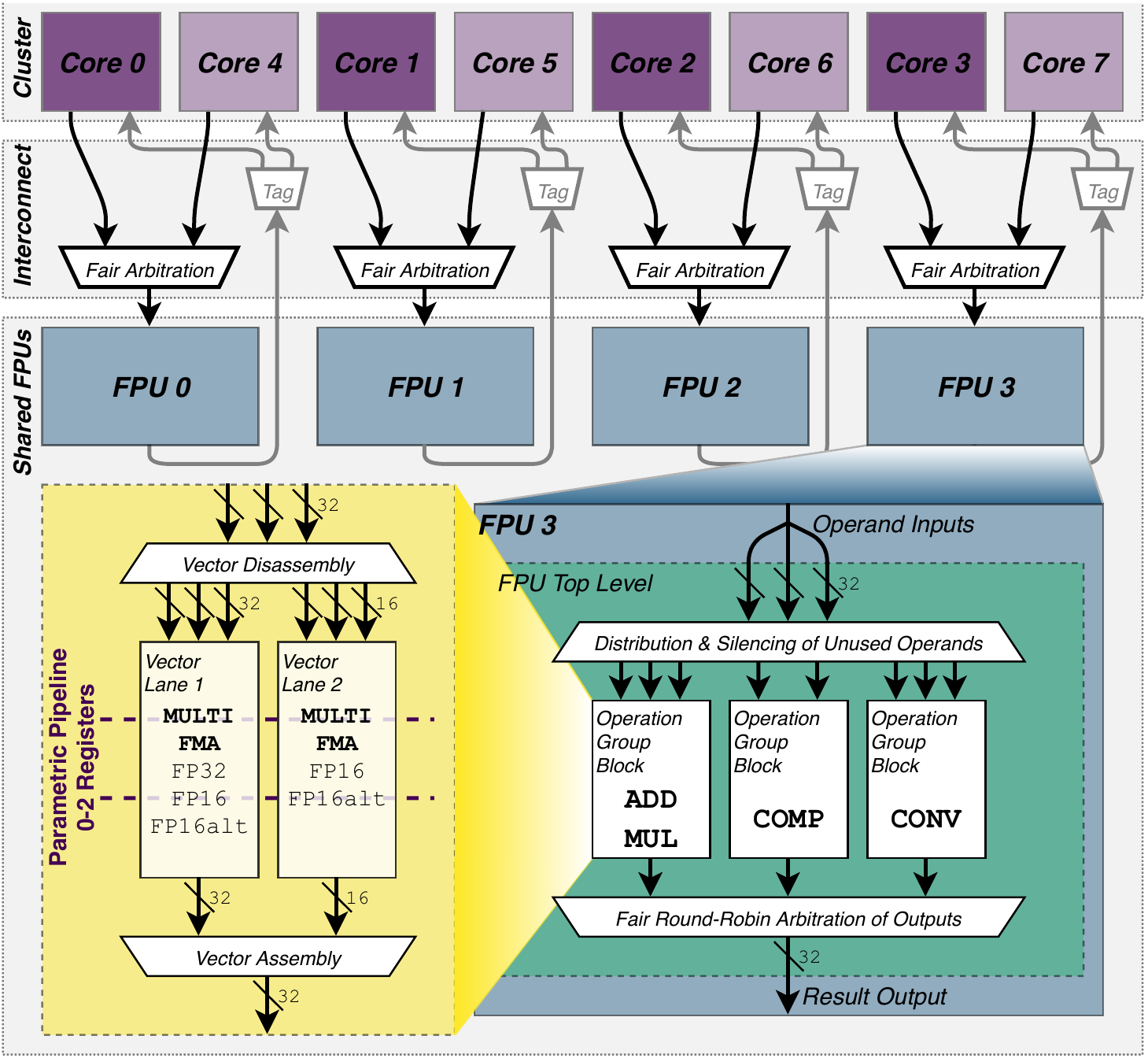}
%\vspace{-6mm}
\caption{FPU sharing for the 8-core, 4-FPU configuration.}
\vspace{-3mm}
\label{fig:fpu_arch}
\end{figure}
The cluster exploits configurations of FPnew \cite{mach2020fpnew} as FPU instances in our evaluation.
FPnew is a parametric FPU architecture that supports multiple FP formats, SIMD vectors, and the insertion of any number of pipeline stages.
\figref{fig:fpu_arch}~(bottom) shows an architectural overview of a single shared FPU instance.
The IP supports the standard IEEE formats, \emph{binary32} (float) and \emph{binary16} (float16), as well as \emph{bfloat16}.
Some operations such as Multiplication and Fused Multiply-Add (FMA) can also be performed as multi-format operations, taking the product of two 16-bit operands but returning a 32-bit single-precision result.
Such multi-format operations are helpful in many near-sensor data analytics applications accumulating data in a higher-precision variable to avoid overflows or losses of precision.
To make full use of the 32-bit data path, we enable packed-SIMD operations for the 16-bit types, boosting the execution performance when using 16-bit data types.
Division and square root operations are disabled in the FPU instances as these operations reside in stand-alone blocks (DIV-SQRT), which are shared separately.
The DIV-SQRT units feature a fixed latency of 11, 7, and 6 cycles for \emph{float}, \emph{float16}, and \emph{bfloat16}, respectively.
Moreover, since DIV-SQRT is designed as an iterative block, back to back pipelined operations are not possible when using these units.

The individual FPU instances are linked to one or more cores through a logarithmic tree interconnect, allowing to share one FPU among multiple cores in a fully transparent way from a software perspective.
On the core side, the interface of the interconnect replaces the FPU in the execution stage, mimicking a core-private unit.
The FPU instances connect to the cores through an auxiliary processing unit (APU) interface, featuring a \emph{ready}/\emph{valid} handshake and support tagging of all in-flight operations, requiring no modification to be shared.

In the proposed design, we employ a partial interconnect with a static mapping of FPUs to cores, such that a core (or a group of cores) will always access the same physical FPU instance.
It arbitrates cases of simultaneous accesses to the FPU by using a fair round-robin policy and propagating the ready signal to only one core, stalling other competing cores.
As such, the fact that FPUs are shared is transparent to both the core and FPU instances.
Moreover, we use a connection scheme with interleaved allocation to reduce access contentions on the FPUs in unbalanced workloads.
For example, in a configuration featuring eight cores and four FPUs, units 0, 1, 2, 3 are shared among cores 0\,\&\,4, 1\,\&\,5, 2\,\&\,6, and 3\,\&\,7, respectively, as shown in \figref{fig:fpu_arch}~(top).
This approach reduces the area and timing overhead compared to a monolithic, fully connected crossbar, which puts significant pressure on the paths from the cores to the first pipeline stage of the FPU, severely limiting the cluster's operating frequency and jeopardizing energy efficiency.
Moreover, it provides an almost optimal allocation (only up to 1\% overhead in performance has been measured against a fully connected crossbar) avoiding contentions on the shared units also when the number of workers in parallel sections is smaller than the number of cores.

In the remainder of the paper, we present a design space exploration of the proposed transprecision cluster, modifying the key configuration parameters presented previously in this section, namely the pipeline stages and sharing factor.
The rationale for the former lies in the fact that in most near sensor-data analytics applications, the density of FPU instructions is smaller than 50\%, hence employing a private, per core FPU may form a bottleneck for area and energy.
On the other hand, the pipelining of the FPU provides a powerful knob to tune the performance and energy efficiency of the transprecision cluster.
If the number of cores and FPUs is equal (1/1 sharing factor), the system effectively degenerates into a core-private scenario, and the interconnect disappears from the design.
In all the considered configurations, a single DIV-SQRT unit is shared among all cores.
Finally, the proposed exploration involves designs of 8-core and 16-core clusters with supply voltages ranging from 0.65 V to 0.8 V to explore the whole design space in between energy-efficient and high-performance solutions.

\subsection{Implementation}
\label{sec:cluster_impl}
\begin{table}[t]
\caption{Description of the architectural configurations of the proposed transprecision cluster that compose the design space. Cluster (8-16-cores), FP units (2-16), and pipeline stages (0-2).}\label{tab:tab_design}
\vspace{-2mm}
\centering 
%\scalebox{.70}[.75]{
\begin{tabular}{rccc} 
\hline 
Mnemonic &Cluster &FP units &Pipeline Stages \\
\hline 
8c2f0p &8-cores   &2   &0   \\ 
\hline
8c2f1p &8-cores   &2   &1   \\ 
\hline
8c2f2p &8-cores   &2   &2   \\ 
\hline
8c4f0p  &8-cores     &4   &0    \\ 
\hline
8c4f1p  &8-cores   &4   &1    \\ 
\hline
8c4f2p  &8-cores   &4  &2   \\ 
\hline
8c8f0p &8-cores    &8   &0  \\ 
\hline
8c8f1p &8-cores    &8   &1  \\ 
\hline
8c8f2p &8-cores    &8   &2   \\ 
\hline
16c4f0p  &16-cores  &4   &0   \\ 
\hline
16c4f1p   &16-cores  &4   &1    \\ 
\hline
16c4f2p   &16-cores  &4   &2    \\ 
\hline
16c8f0p   &16-cores  &8   &0    \\ 
\hline
16c8f1p &16-cores   &8   &1   \\ 
\hline
16c8f2p &16-cores  &8   &2    \\ 
\hline 
16c16f0p  &16-cores  &16   &0    \\ 
\hline
16c16f1p &16-cores  &16   &1    \\ 
\hline
16c16f2p &16-cores   &16   &2   \\ 
\hline
\end{tabular}
%} 
%\vspace{-2mm}
\end{table}
This section presents the physical implementation results and related explorations of the proposed cluster.
Table~\ref{tab:tab_design} describes the 18 different configurations, given by the combination of the three architectural parameters (number of cores, number of FP units, and number of FPU pipeline stages), as described in the previous section.
The various configurations of the clusters have been synthesized using Synopsys Design Compiler 2019.12, using LVT libraries from 22nm FDX technology from Global Foundries.
Physical implementation has been performed with Cadence Innovus v19.10-p002\_1, using both 0.65 V near-threshold (NT) and 0.8 V super-threshold (ST) corners.
We considered all permutations of operating conditions for signoff: fast and slow process transistors, $125^\circ$C and $-40^\circ$C temperatures, $\pm$10\% of the voltage supply, as well as optimistic and pessimistic parasitics.
Power analysis has been performed with Synopsys PrimeTime 2019.12 using the nominal corners at 0.65 V and 0.8 V, extracting \emph{value change dump} (VCD) traces through parasitic-annotated post-layout simulation of a 32-bit floating-point matrix multiplication performed using Mentor Modelsim 2008.06.
Each configuration has been synthesized and implemented at its maximum operating frequency.
In contrast, power consumption has been analyzed at the same operating frequency for all configurations (100 MHz) to guarantee a fair comparison.

%
%For shortness, the mnemonics chosen to identify each configuration unambiguously is composed of 3 parts. The first characters represent the number of cores inside the cluster. The following group of characters indicates the number of FPUs (2, 4, 8, or 16). The sharing factor can be derived consequently; e.g., 16c8f describes an architecture with 16 cores and 8 FPUs, thus the sharing factor is 1/2 (1 FPU every 2 cores). The last characters represent the number of pipeline stages (e.g., \emph{2p} for two pipeline stages).
%

\begin{figure}[t]
\begin{center}
\includegraphics[width=\columnwidth]{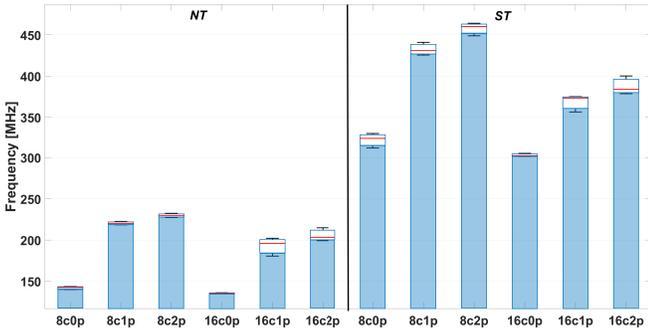}
%\vspace{-6mm}
\caption{Minimum, maximum, and median values of the frequencies for all the configurations of the transprecision cluster, divided in NT and ST voltage corners.}
%\vspace{-3mm}
\label{fig:freq}
\end{center}
\end{figure}

\begin{figure}[t]
\begin{center}
\includegraphics[width=\columnwidth]{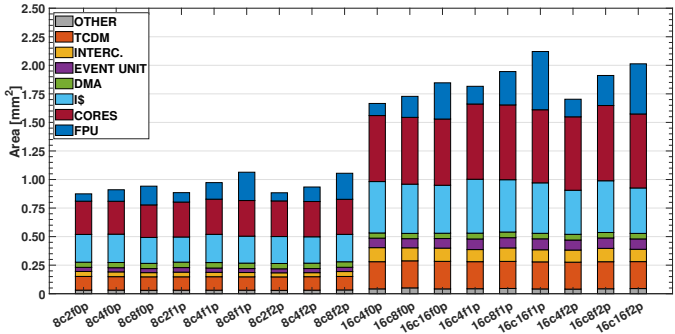}
%\vspace{-6mm}
\caption{Total area of all the configurations in the design space of the transprecision cluster.}
%\vspace{-3mm}
\label{fig:area}
\end{center}
\end{figure}

\begin{figure}[t]
\begin{center}
\includegraphics[width=1\columnwidth]{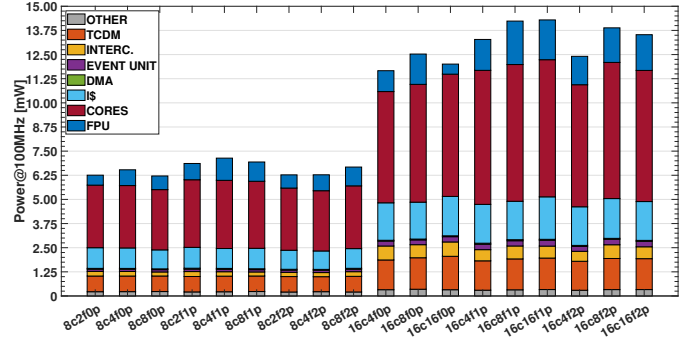}
\caption{Total power consumption (at 100 MHz) of all the configurations in the design space of the transprecision cluster. }
%\vspace{-3mm}
\label{fig:power}
\end{center}
\end{figure}

Fig.~\ref{fig:freq}, Fig.~\ref{fig:area} and Fig.~\ref{fig:power} show the frequency, the area, and the power consumption of the cluster configurations analyzed in this work at 100 MHz.
In Fig.~\ref{fig:freq}, we report the minimum, maximum, and median values of the frequencies obtained varying the number of FPUs.
When considering single-cycle latency FPUs, we note that the entire system's operating frequency suffers profoundly.
The long paths starting from the ID/EX registers of the core towards the FPUs and then back to the EX/WB registers form a considerable bottleneck for operating frequency.
On the other hand, the absence of pipeline registers makes this solution quite small and low-power.
When moving to single-stage pipeline solutions, we note a very significant increase in the operating frequency when using NT cells (almost 50\%).
In contrast, the performance increase using ST cells is more limited since the design already hits a structurally critical path from the TCDM SRAMs (featuring wide-voltage range but low-performance in ST) to the core through the logarithmic interconnect.
In all configurations featuring one pipeline stage, we can observe an increase of power and area, due to the extra overhead of the additional pipeline stage.
When adding a second pipeline stage to the FPUs, we can see another slight increase of frequency in all configurations.
In these configurations, we also encounter structurally critical paths using NT, through control paths of the interconnect to the instruction cache.
With two pipeline stages, although the area increases for all configurations, the power consumption tends to decrease thanks to the smaller timing pressure on the FPU.

Considering the sharing factor, we note that the impact on frequency caused by the FPU interconnect is negligible and that the area linearly increases when moving from 1/4 to 1/1 sharing, for all configurations.
On the other hand, when moving from 1/4 to 1/2 sharing factor, the power increases significantly due to the high utilization of the units.
When moving from 1/2 to 1/1 sharing, we note that the power consumption decreases in almost all cases. This effect occurs because even if we consider a highly intensive benchmark (e.g., matrix multiplication), the FP intensity around 50\% leads to underutilization of the available resources, causing smaller power consumption.
Additionally, the 1/1 configuration removes the interconnect, which relaxes the paths through the FPU, leading to smaller power consumption.
Finally, if we consider the scaling of the number of cores, we can notice that most of the power components scale linearly with the number of cores (i.e., core power, TCDM power, and FPU power).
On the other hand, other components such as the interconnect and the instruction cache scale superlinearly, indicating a smaller efficiency for the 16-core configuration.
Moreover, the operating frequency of the 16-core cluster decreases compared to the one using eight cores. This effect is due to the longer path through the interconnects. Finally, we can notice that the area increases less than linearly due to some blocks not being duplicated, such as the DMA, the event unit, and the shared banks of the I\$.

 \section{Programming model and compilation toolchain}
\label{sec:software}
The full exploitation of the transprecision cluster proposed in this work requires the support of a comprehensive software ecosystem, including a parallel programming model, vectorization techniques, and compiler support.

% HAL runtime and parallel programming support
The architectural template of the transprecision cluster promotes a Single-Program Multiple-Data (SPMD) parallel paradigm. This paradigm is supported by a Hardware Abstraction Layer (HAL), which allows minimal access to the platform features. The HAL provides information such as the identifier of the core that can be used to organize the parallel workload for both data and task parallelism. 
In this programming model, all the cores of the cluster follow the same execution flow, unless the programmer explicitly indicates that a specific region should be executed by a subset of the cores, splitting the workload among the cores running concurrently on different data. Inter-core synchronization barriers are explicitly indicated to ensure the correctness of the results. 
Our architecture features dedicated hardware support that allows optimizing synchronization construct like barriers or critical sections. 
The HAL layer provides the basic primitives to support high-level parallel programming models such as OpenMP. These models can provide a more intuitive interface at the cost of higher overhead; however, our exploration is focused on finding the maximum performance that we can obtain from real applications without considering a multi-layer software stack.

% Vectorization
To provide compiler support, We have extended the RISC-V GCC toolchain with new data types (float16, bfloat16) and support for manual and automatic vectorization using packed-SIMD instructions.
% Manual vectorization
Programmers can explicitly use vector data types through a \texttt{typedef} declaration coupled with a \texttt{vector\_size} attribute; the compiler automatically lowers standard arithmetic and comparison operations involving these types into their vector counterpart.
The ISA extension also includes cast-and-pack operations that convert two scalar single-precision operands and insert them into two adjacent entries of a packed vector in the destination register.
These operations aim at removing the bottleneck of ``convert scalars and assemble vectors'' operations that could seriously compromise performance and energy efficiency of transprecision computing techniques \cite{tagliavini2018transprecision}.
A set of compiler intrinsics provides access to cast-and-pack operations.
% Automatic vectorization
The automatic vectorization pass of GCC operates on the middle-end intermediate representation \footnote{\url{https://www.gnu.org/software/gcc/projects/tree-ssa/vectorization.html}}.
This pass analyzes the loops to replace the scalar operations with the vectorial ones reducing the loop trip-count by the vectorization factor.
We have also extended the standard GCC auto-vectorizer to use cast-and-pack operations. The original version only recognizes patterns involving multiple vector types with different widths using unrolling and vector-to-vector casts (i.e., cast-and-pack semantic was not supported).

In this work, we further extend the compiler back-end to support a parametric number of FPU pipeline stages.
This parameter has a substantial impact on the instruction scheduling algorithm: imprecise modeling of the FPU instruction latency may introduce stalls due to data dependencies with the result.
We have modified the FPU pipeline description to include the hardware functional units and introduce a command-line option to specify the number of stages in the target configuration.
Based on this option, the model specifies different latency and reservation delay for the functional units involved in FP operations.
Finally, we have a set of platform-specific parameters for the instruction scheduling algorithm.
This algorithm uses a heuristic function to estimate the relative costs of operations; this value enables the choice of the best assembly sequence in case of multiple alternatives in the lowering process.
 \section{Experimental Results}
\label{sec:experiments}

\subsection{Experimental Set-up}
The experiments have been performed on a hardware emulator implemented on a Xilinx UltraScale+ VCU118 FPGA board\footnote{\url{https://github.com/pulp-platform/pulp/tree/master/fpga/pulpissimo-zcu104}}.
The emulation on the FPGA provides cycle-accurate results, with a significant speed-up of the experiments compared to an RTL-equivalent simulation.

A set of non-intrusive per-core performance counters included in the hardware design record the number of executed instructions and cycles spent in different states (total, active, L2/TCDM memory stalls, TCDM contention, FPU stall, FPU contention, FPU write-back stall, instruction cache miss).
We have generated all the bitstreams for all the configurations reported in Table~\ref{tab:tab_design} and, after loading a bitstream on the FPGA, we load and run application binaries using OpenOCD and GDB interfaces.
%enables to load the bitstream corresponding to the current configuration on the FPGA.
The same interface is used to load a program binary in the L2 memory, start the program execution, and finally read the performance counters from an emulated terminal.

The values of power consumption used to calculate the efficiency have been derived from an annotated post-layout simulation, as described in Section~\ref{sec:cluster_impl}.
\begin{table}[t]
\caption{Main application domains (Domains), FP intensity (FP I.), and memory intensity (M. I.) for scalar and vector variants of the benchmarks. %\textbf{CONTROLLARE NUMERI. TODO FABIO Are the numbers ok? Can we remove this label?}
}\label{tab:tab_app}
\vspace{-2mm}
\centering 
%\scalebox{.70}[.75]{
\begin{tabular}{|l|c|c|c|c|c|} 
\hline 
 & & \multicolumn{2}{c|}{\textbf{Scalar}} & \multicolumn{2}{c|}{\textbf{Vector}} \\
\hline 
\textbf{Apps} &\textbf{Domains}      &\textbf{FP I.} &\textbf{M. I.} &\textbf{FP I.} &\textbf{M. I.}\\
\hline 
CONV &Audio, Image, ExG        &0.33	&0.67	&0.28	&0.29\\ 
\hline
DWT &Audio, Image, ExG        &0.29	&0.59	&0.21	&0.57\\ \hline
FFT &Audio, Image, ExG         &0.32	&0.52	&0.26	&0.38\\  
\hline
FIR   &Audio, Image, ExG      &0.32	&0.65	&0.32	&0.48\\ \hline
IIR &Audio, Image, ExG      &0.19	&0.55	&0.17	&0.33\\ 
\hline
KMEANS &ExG                   &0.55	&0.36	&0.44	&0.30\\ 
\hline
MATMUL &Audio, Image, ExG      &0.28	&0.58	&0.27	&0.41\\  
\hline
SVM &ExG                      &0.27	&0.53	&0.21	&0.52\\  
\hline
\end{tabular}
%} 
%\vspace{-2mm}
\end{table}

\subsection{Benchmarks}
\label{sec:benchmarks}
To evaluate the different configurations of the proposed transprecision cluster architecture, we analyzed eight benchmarks commonly used in the near-sensor processing applications for filtering, feature extraction, classification, and basic linear algebra functions. 
Table \ref{tab:tab_app} illustrates the target benchmarks associated with their domains (i.e., audio processing, image processing, ExG biosignal processing).
%
% Please add the following required packages to your document preamble:
% \usepackage{multirow}
% \usepackage[table.xcdraw]{xcolor}
% If you use beamer only pass "xcolor=table" option. i.e. \documentclass[xcolor=table]{beamer}
\begin{table*}[t]
\caption{Performance [$Gflop/s$], energy efficiency [$Gflop/s/W$], and area efficiency [$Gflop/s/mm^{2}$] executing the benchmarks on the 8-cores configurations. Performance and area efficiency are computed at 0.8 $V$, energy efficiency at 0.65 $V$. The last group of lines reports normalized average values. A box around the metric value highlights the best configuration for each benchmark.
%\textbf{TODO FABIO Ue . instead of . as decimal separator!!! CHECK ALL THE TABLES AND ALL THE FIGURES!!!}
}
\label{tab:results_8_cores}
\resizebox{\textwidth}{!}{
\begin{tabular}{cc|c|c|c|c|c|c|c|c|c|c|c|c|c|c|c|c|c|c|}
\cline{3-20}
 & \multicolumn{1}{c|}{} & \multicolumn{9}{c|}{\textbf{Scalar}} & \multicolumn{9}{c|}{\textbf{Vector}} \\ \cline{3-20} 
\multicolumn{2}{c}{} & \multicolumn{1}{|c|}{\textbf{8c2f0p}} & \multicolumn{1}{c|}{\textbf{8c2f1p}} & \multicolumn{1}{c|}{\textbf{8c2f2p}} & \multicolumn{1}{c|}{\textbf{8c4f0p}} & \multicolumn{1}{c|}{\textbf{8c4f1p}} & \multicolumn{1}{c|}{\textbf{8c4f2p}} & \multicolumn{1}{c|}{\textbf{8c8f0p}} & \multicolumn{1}{c|}{\textbf{8c8f1p}} & \multicolumn{1}{c|}{\textbf{8c8f2p}} & \multicolumn{1}{c|}{\textbf{8c2f0p}} & \multicolumn{1}{c|}{\textbf{8c2f1p}} & \multicolumn{1}{c|}{\textbf{8c2f2p}} & \multicolumn{1}{c|}{\textbf{8c4f0p}} & \multicolumn{1}{c|}{\textbf{8c4f1p}} & \multicolumn{1}{c|}{\textbf{8c4f2p}} & \multicolumn{1}{c|}{\textbf{8c8f0p}} & \multicolumn{1}{c|}{\textbf{8c8f1p}} & \multicolumn{1}{c|}{\textbf{8c8f2p}} \\ \hline
\multicolumn{1}{|c|}{} & \textbf{PERF} & \cellcolor[HTML]{DDF0E4}1.16 & \cellcolor[HTML]{C6E6D0}1.62 & \cellcolor[HTML]{C0E4CB}1.74 & \cellcolor[HTML]{CFEAD8}1.43 & \cellcolor[HTML]{B6E0C2}1.94 & \cellcolor[HTML]{B4DFC1}1.97 & \cellcolor[HTML]{CBE8D5}1.52 & \cellcolor[HTML]{B0DEBE}2.04 & \cellcolor[HTML]{B4DFC1}1.96 & \cellcolor[HTML]{B7E0C4}1.91 & \cellcolor[HTML]{96D3A7}2.57 & \cellcolor[HTML]{A9DBB7}2.19 & \cellcolor[HTML]{B0DEBE}2.04 & \cellcolor[HTML]{85CC99}2.89 & \cellcolor[HTML]{A0D7B0}2.36 & \cellcolor[HTML]{A2D8B2}2.32 & \cellcolor[HTML]{81CA95}2.98 & \cellcolor[HTML]{A1D7B0}2.35 \\ \cline{2-20}
\multicolumn{1}{|c|}{} & \textbf{E. EFF} & \cellcolor[HTML]{FBCDCF}72 & \cellcolor[HTML]{FBD3D6}66 & \cellcolor[HTML]{FBCDCF}72 & \cellcolor[HTML]{FBC2C5}82 & \cellcolor[HTML]{FBC8CB}76 & \cellcolor[HTML]{FBC0C3}83 & \cellcolor[HTML]{FBB7BA}91 & \cellcolor[HTML]{FBC1C3}83 & \cellcolor[HTML]{FBC3C6}81 & \cellcolor[HTML]{FA9A9C}119 & \cellcolor[HTML]{FAA8AB}105 & \cellcolor[HTML]{FBB8BB}91 & \cellcolor[HTML]{FA9C9E}117 & \cellcolor[HTML]{FA9FA2}113 & \cellcolor[HTML]{FAAEB1}100 & \cellcolor[HTML]{F98385}139 & \cellcolor[HTML]{FA9799}121 & \cellcolor[HTML]{FAB1B4}97 \\ \cline{2-20} 
\multicolumn{1}{|c|}{\multirow{-3}{*}{\rotatebox[origin=c]{90}{\textbf{CONV}}}} & \textbf{A. EFF} & \cellcolor[HTML]{C5DBF3}1.5 & \cellcolor[HTML]{B1CFEF}1.8 & \cellcolor[HTML]{A8C9ED}2.0 & \cellcolor[HTML]{B3D0F0}1.8 & \cellcolor[HTML]{A6C8ED}\framebox{2.0} & \cellcolor[HTML]{9FC3EB}2.1 & \cellcolor[HTML]{C0D8F2}1.6 & \cellcolor[HTML]{ABCBEE}1.9 & \cellcolor[HTML]{AFCDEF}1.9 & \cellcolor[HTML]{83B2E6}2.5 & \cellcolor[HTML]{6AA2E0}2.9 & \cellcolor[HTML]{86B4E6}2.5 & \cellcolor[HTML]{80B0E5}2.6 & \cellcolor[HTML]{659FDF}\framebox{3.0} & \cellcolor[HTML]{83B2E6}2.5 & \cellcolor[HTML]{87B5E7}2.5 & \cellcolor[HTML]{70A6E2}2.8 & \cellcolor[HTML]{97BEEA}2.2 \\ \hline 
\multicolumn{1}{|c|}{} & \textbf{PERF} & \cellcolor[HTML]{FCFCFF}0.54 & \cellcolor[HTML]{F2F8F7}0.73 & \cellcolor[HTML]{F1F8F5}0.77 & \cellcolor[HTML]{F4F9F8}0.70 & \cellcolor[HTML]{EBF6F1}0.87 & \cellcolor[HTML]{ECF6F1}0.86 & \cellcolor[HTML]{F1F8F6}0.75 & \cellcolor[HTML]{E7F4ED}0.95 & \cellcolor[HTML]{ECF6F2}0.85 & \cellcolor[HTML]{EEF6F3}0.83 & \cellcolor[HTML]{DFF0E6}1.12 & \cellcolor[HTML]{DCEFE3}1.18 & \cellcolor[HTML]{EBF5F0}0.89 & \cellcolor[HTML]{DCEFE4}1.17 & \cellcolor[HTML]{DBEFE3}1.19 & \cellcolor[HTML]{E9F5EF}0.92 & \cellcolor[HTML]{DAEFE2}1.21 & \cellcolor[HTML]{DDF0E4}1.16 \\ \cline{2-20} 
\multicolumn{1}{|c|}{} & \textbf{E. EFF} & \cellcolor[HTML]{FCF7FA}33 & \cellcolor[HTML]{FCFBFE}30 & \cellcolor[HTML]{FCF9FC}32 & \cellcolor[HTML]{FCF0F3}40 & \cellcolor[HTML]{FCF6F9}34 & \cellcolor[HTML]{FCF4F7}36 & \cellcolor[HTML]{FCEAED}45 & \cellcolor[HTML]{FCF1F4}39 & \cellcolor[HTML]{FCF5F8}35 & \cellcolor[HTML]{FCE3E6}51 & \cellcolor[HTML]{FCE9EC}46 & \cellcolor[HTML]{FCE6E9}49 & \cellcolor[HTML]{FCE4E7}51 & \cellcolor[HTML]{FCE9EC}46 & \cellcolor[HTML]{FCE4E7}50 & \cellcolor[HTML]{FCDFE2}55 & \cellcolor[HTML]{FCE6E8}49 & \cellcolor[HTML]{FCE7EA}48 \\ \cline{2-20} 
\multicolumn{1}{|c|}{\multirow{-3}{*}{\rotatebox[origin=c]{90}{\textbf{DWT}}}} & \textbf{A. EFF} & \cellcolor[HTML]{FCFDFF}0.7 & \cellcolor[HTML]{F4F8FD}0.8 & \cellcolor[HTML]{F1F7FD}0.9 & \cellcolor[HTML]{F1F6FC}0.9 & \cellcolor[HTML]{EFF6FC}\framebox{0.9} & \cellcolor[HTML]{EEF5FC}0.9 & \cellcolor[HTML]{F6FAFE}0.8 & \cellcolor[HTML]{F0F6FC}0.9 & \cellcolor[HTML]{F5F9FD}0.8 & \cellcolor[HTML]{E2EDF9}1.1 & \cellcolor[HTML]{D7E6F7}1.3 & \cellcolor[HTML]{D2E3F6}1.3 & \cellcolor[HTML]{E1ECF9}1.1 & \cellcolor[HTML]{DBE9F8}\framebox{1.2} & \cellcolor[HTML]{D6E6F7}1.3 & \cellcolor[HTML]{EAF2FB}1.0 & \cellcolor[HTML]{DFEBF9}1.1 & \cellcolor[HTML]{E2EDF9}1.1 \\ \hline
\multicolumn{1}{|c|}{} & \textbf{PERF} & \cellcolor[HTML]{F6FAFA}0.67 & \cellcolor[HTML]{EAF5EF}0.91 & \cellcolor[HTML]{E7F4ED}0.97 & \cellcolor[HTML]{E9F5EF}0.92 & \cellcolor[HTML]{DAEEE1}1.23 & \cellcolor[HTML]{DAEFE2}1.21 & \cellcolor[HTML]{E4F3EA}1.02 & \cellcolor[HTML]{D2EBDB}1.37 & \cellcolor[HTML]{D7EDDF}1.27 & \cellcolor[HTML]{DAEFE2}1.21 & \cellcolor[HTML]{C9E8D3}1.56 & \cellcolor[HTML]{C9E8D3}1.54 & \cellcolor[HTML]{CCE9D6}1.49 & \cellcolor[HTML]{BBE2C7}1.83 & \cellcolor[HTML]{C3E5CE}1.66 & \cellcolor[HTML]{C7E7D1}1.60 & \cellcolor[HTML]{B4DFC1}1.98 & \cellcolor[HTML]{C5E6D0}1.63 \\ \cline{2-20} 
\multicolumn{1}{|c|}{} & \textbf{E. EFF} & \cellcolor[HTML]{FCEEF1}42 & \cellcolor[HTML]{FCF3F6}37 & \cellcolor[HTML]{FCF0F2}40 & \cellcolor[HTML]{FCE2E5}52 & \cellcolor[HTML]{FCE7EA}48 & \cellcolor[HTML]{FCE4E6}51 & \cellcolor[HTML]{FCD8DB}61 & \cellcolor[HTML]{FCDFE1}56 & \cellcolor[HTML]{FCE2E5}52 & \cellcolor[HTML]{FBC9CC}75 & \cellcolor[HTML]{FBD6D8}64 & \cellcolor[HTML]{FBD5D8}64 & \cellcolor[HTML]{FBBEC1}85 & \cellcolor[HTML]{FBCDD0}72 & \cellcolor[HTML]{FBCED1}70 & \cellcolor[HTML]{FAB2B5}96 & \cellcolor[HTML]{FBC4C6}80 & \cellcolor[HTML]{FBD2D5}67 \\ \cline{2-20} 
\multicolumn{1}{|c|}{\multirow{-3}{*}{\rotatebox[origin=c]{90}{\textbf{FFT}}}} & \textbf{A. EFF} & \cellcolor[HTML]{F0F6FC}0.9 & \cellcolor[HTML]{E7F0FA}1.0 & \cellcolor[HTML]{E2EDF9}1.1 & \cellcolor[HTML]{DEEBF9}1.2 & \cellcolor[HTML]{D7E6F7}\framebox{1.3} & \cellcolor[HTML]{D5E5F7}1.3 & \cellcolor[HTML]{E3EEFA}1.1 & \cellcolor[HTML]{D5E5F7}1.3 & \cellcolor[HTML]{DBE9F8}1.2 & \cellcolor[HTML]{C1D9F3}1.6 & \cellcolor[HTML]{B6D2F0}1.8 & \cellcolor[HTML]{B7D2F0}1.7 & \cellcolor[HTML]{AECDEF}1.9 & \cellcolor[HTML]{AECDEF}\framebox{1.9} & \cellcolor[HTML]{B4D1F0}1.8 & \cellcolor[HTML]{BAD4F1}1.7 & \cellcolor[HTML]{AFCEEF}1.9 & \cellcolor[HTML]{C4DAF3}1.5 \\ \hline
\multicolumn{1}{|c|}{} & \textbf{PERF} & \cellcolor[HTML]{DAEFE2}1.21 & \cellcolor[HTML]{CAE8D4}1.54 & \cellcolor[HTML]{D1EBDA}1.40 & \cellcolor[HTML]{CCE9D6}1.49 & \cellcolor[HTML]{BFE4CA}1.76 & \cellcolor[HTML]{CDE9D6}1.48 & \cellcolor[HTML]{C6E6D0}1.62 & \cellcolor[HTML]{B9E1C5}1.88 & \cellcolor[HTML]{CDE9D7}1.47 & \cellcolor[HTML]{A6DAB5}2.24 & \cellcolor[HTML]{7EC993}3.03 & \cellcolor[HTML]{8CCF9E}2.76 & \cellcolor[HTML]{97D4A8}2.54 & \cellcolor[HTML]{6DC284}3.38 & \cellcolor[HTML]{87CD9A}2.86 & \cellcolor[HTML]{8FD0A1}2.70 & \cellcolor[HTML]{63BE7B}3.57 & \cellcolor[HTML]{8BCE9D}2.79 \\ \cline{2-20} 
\multicolumn{1}{|c|}{} & \textbf{E. EFF} & \cellcolor[HTML]{FBC9CC}75 & \cellcolor[HTML]{FBD7D9}63 & \cellcolor[HTML]{FCDCDF}58 & \cellcolor[HTML]{FBBEC1}85 & \cellcolor[HTML]{FBD0D3}69 & \cellcolor[HTML]{FBD7DA}63 & \cellcolor[HTML]{FAB1B3}97 & \cellcolor[HTML]{FBC8CB}76 & \cellcolor[HTML]{FCD9DC}61 & \cellcolor[HTML]{F98385}139 & \cellcolor[HTML]{FA9396}124 & \cellcolor[HTML]{FA9EA0}114 & \cellcolor[HTML]{F97D7F}145 & \cellcolor[HTML]{F98A8D}132 & \cellcolor[HTML]{FA9799}121 & \cellcolor[HTML]{F8696B}162 & \cellcolor[HTML]{F97D7F}145 & \cellcolor[HTML]{FA9EA0}115 \\ \cline{2-20} 
\multicolumn{1}{|c|}{\multirow{-3}{*}{\rotatebox[origin=c]{90}{\textbf{FIR}}}} & \textbf{A. EFF} & \cellcolor[HTML]{C1D9F3}1.6 & \cellcolor[HTML]{B8D3F1}1.7 & \cellcolor[HTML]{C2D9F3}1.6 & \cellcolor[HTML]{AECDEF}1.9 & \cellcolor[HTML]{B3D0F0}\framebox{1.8} & \cellcolor[HTML]{C1D9F3}1.6 & \cellcolor[HTML]{B9D4F1}1.7 & \cellcolor[HTML]{B6D2F0}1.8 & \cellcolor[HTML]{CEE1F5}1.4 & \cellcolor[HTML]{66A0E0}3.0 & \cellcolor[HTML]{478DD9}3.4 & \cellcolor[HTML]{5B99DD}3.1 & \cellcolor[HTML]{5696DC}3.2 & \cellcolor[HTML]{438AD8}\framebox{3.5} & \cellcolor[HTML]{5F9CDE}3.1 & \cellcolor[HTML]{6CA4E1}2.9 & \cellcolor[HTML]{4C90DA}3.3 & \cellcolor[HTML]{7BADE4}2.6 \\ \hline
\multicolumn{1}{|c|}{} & \textbf{PERF} & \cellcolor[HTML]{F9FBFC}0.61 & \cellcolor[HTML]{EEF7F3}0.82 & \cellcolor[HTML]{ECF6F1}0.86 & \cellcolor[HTML]{F4F9F8}0.70 & \cellcolor[HTML]{EAF5F0}0.90 & \cellcolor[HTML]{E9F5EF}0.91 & \cellcolor[HTML]{F2F8F6}0.74 & \cellcolor[HTML]{E8F4EE}0.94 & \cellcolor[HTML]{EAF5EF}0.91 & \cellcolor[HTML]{E2F2E9}1.06 & \cellcolor[HTML]{D1EBDA}1.40 & \cellcolor[HTML]{CEEAD7}1.46 & \cellcolor[HTML]{DDF0E5}1.15 & \cellcolor[HTML]{CCE9D6}1.49 & \cellcolor[HTML]{CCE9D6}1.49 & \cellcolor[HTML]{DBEFE3}1.19 & \cellcolor[HTML]{C9E8D3}1.55 & \cellcolor[HTML]{CDE9D6}1.48 \\ \cline{2-20} 
\multicolumn{1}{|c|}{} & \textbf{E. EFF} & \cellcolor[HTML]{FCF2F5}38 & \cellcolor[HTML]{FCF7FA}33 & \cellcolor[HTML]{FCF5F8}35 & \cellcolor[HTML]{FCF0F2}40 & \cellcolor[HTML]{FCF5F8}35 & \cellcolor[HTML]{FCF1F4}39 & \cellcolor[HTML]{FCEBEE}45 & \cellcolor[HTML]{FCF2F5}38 & \cellcolor[HTML]{FCF3F6}37 & \cellcolor[HTML]{FBD4D6}66 & \cellcolor[HTML]{FCDCDF}57 & \cellcolor[HTML]{FCD9DC}60 & \cellcolor[HTML]{FBD4D6}65 & \cellcolor[HTML]{FCDBDE}58 & \cellcolor[HTML]{FBD6D9}63 & \cellcolor[HTML]{FBCDD0}72 & \cellcolor[HTML]{FBD7DA}63 & \cellcolor[HTML]{FCD9DC}61 \\ \cline{2-20} 
\multicolumn{1}{|c|}{\multirow{-3}{*}{\rotatebox[origin=c]{90}{\textbf{IIR}}}} & \textbf{A. EFF} & \cellcolor[HTML]{F6FAFE}0.8 & \cellcolor[HTML]{EEF4FC}0.9 & \cellcolor[HTML]{EBF3FB}1.0 & \cellcolor[HTML]{F0F6FC}0.9 & \cellcolor[HTML]{EEF4FC}\framebox{0.9} & \cellcolor[HTML]{EAF2FB}1.0 & \cellcolor[HTML]{F7FAFE}0.8 & \cellcolor[HTML]{F1F6FC}0.9 & \cellcolor[HTML]{F2F7FD}0.9 & \cellcolor[HTML]{CEE1F5}1.4 & \cellcolor[HTML]{C1D9F3}1.6 & \cellcolor[HTML]{BDD6F2}1.6 & \cellcolor[HTML]{CBDFF5}1.4 & \cellcolor[HTML]{C5DBF3}\framebox{1.5} & \cellcolor[HTML]{C1D9F3}1.6 & \cellcolor[HTML]{D7E6F7}1.3 & \cellcolor[HTML]{CADEF4}1.5 & \cellcolor[HTML]{CEE1F5}1.4 \\ \hline
\multicolumn{1}{|c|}{} & \textbf{PERF} & \cellcolor[HTML]{F1F8F6}0.75 & \cellcolor[HTML]{E4F3EA}1.02 & \cellcolor[HTML]{E2F2E9}1.05 & \cellcolor[HTML]{DDF0E5}1.15 & \cellcolor[HTML]{CCE9D6}1.49 & \cellcolor[HTML]{D6EDDE}1.30 & \cellcolor[HTML]{D4ECDC}1.34 & \cellcolor[HTML]{C2E5CD}1.68 & \cellcolor[HTML]{D6EDDE}1.30 & \cellcolor[HTML]{DAEFE2}1.21 & \cellcolor[HTML]{C4E6CF}1.64 & \cellcolor[HTML]{C1E4CC}1.72 & \cellcolor[HTML]{C3E5CE}1.68 & \cellcolor[HTML]{ACDCBA}2.13 & \cellcolor[HTML]{B0DDBD}2.06 & \cellcolor[HTML]{B8E1C5}1.88 & \cellcolor[HTML]{A2D8B1}2.33 & \cellcolor[HTML]{AEDDBB}2.10 \\ \cline{2-20} 
\multicolumn{1}{|c|}{} & \textbf{E. EFF} & \cellcolor[HTML]{FCE8EB}47 & \cellcolor[HTML]{FCEEF1}42 & \cellcolor[HTML]{FCECEF}43 & \cellcolor[HTML]{FBD4D6}66 & \cellcolor[HTML]{FCDCDE}58 & \cellcolor[HTML]{FCDFE2}55 & \cellcolor[HTML]{FBC3C6}80 & \cellcolor[HTML]{FBD1D3}68 & \cellcolor[HTML]{FCE1E4}53 & \cellcolor[HTML]{FBC9CC}75 & \cellcolor[HTML]{FBD2D4}67 & \cellcolor[HTML]{FBCDD0}71 & \cellcolor[HTML]{FAB2B5}96 & \cellcolor[HTML]{FBC0C3}83 & \cellcolor[HTML]{FBBCBF}87 & \cellcolor[HTML]{FA9FA2}113 & \cellcolor[HTML]{FBB4B7}94 & \cellcolor[HTML]{FBBDC0}86 \\ \cline{2-20} 
\multicolumn{1}{|c|}{\multirow{-3}{*}{\rotatebox[origin=c]{90}{\textbf{K-M.}}}} & \textbf{A. EFF} & \cellcolor[HTML]{E9F2FB}1.0 & \cellcolor[HTML]{DFEBF9}1.1 & \cellcolor[HTML]{DCE9F8}1.2 & \cellcolor[HTML]{CBDFF5}1.4 & \cellcolor[HTML]{C5DBF3}\framebox{1.5} & \cellcolor[HTML]{CEE1F5}1.4 & \cellcolor[HTML]{CDE0F5}1.4 & \cellcolor[HTML]{C2D9F3}1.6 & \cellcolor[HTML]{D9E8F8}1.2 & \cellcolor[HTML]{C1D9F3}1.6 & \cellcolor[HTML]{AFCEEF}1.9 & \cellcolor[HTML]{A9CAEE}1.9 & \cellcolor[HTML]{9EC3EB}2.1 & \cellcolor[HTML]{99C0EA}\framebox{2.2} & \cellcolor[HTML]{98BFEA}2.2 & \cellcolor[HTML]{A6C8ED}2.0 & \cellcolor[HTML]{9AC0EA}2.2 & \cellcolor[HTML]{A7C8ED}2.0 \\ \hline
\multicolumn{1}{|c|}{} & \textbf{PERF} & \cellcolor[HTML]{E2F2E9}1.06 & \cellcolor[HTML]{CAE8D4}1.54 & \cellcolor[HTML]{CBE9D5}1.51 & \cellcolor[HTML]{D7EDDF}1.28 & \cellcolor[HTML]{C0E4CB}1.74 & \cellcolor[HTML]{C6E6D1}1.61 & \cellcolor[HTML]{D3ECDC}1.35 & \cellcolor[HTML]{BCE2C8}1.81 & \cellcolor[HTML]{C6E6D0}1.61 & \cellcolor[HTML]{ADDCBB}2.10 & \cellcolor[HTML]{8CCF9E}2.77 & \cellcolor[HTML]{90D1A2}2.68 & \cellcolor[HTML]{A0D7B0}2.36 & \cellcolor[HTML]{78C78D}3.16 & \cellcolor[HTML]{8CCF9E}2.77 & \cellcolor[HTML]{9BD5AB}2.46 & \cellcolor[HTML]{70C386}3.32 & \cellcolor[HTML]{8FD0A1}2.71 \\ \cline{2-20} 
\multicolumn{1}{|c|}{} & \textbf{E. EFF} & \cellcolor[HTML]{FBD4D6}66 & \cellcolor[HTML]{FBD7D9}63 & \cellcolor[HTML]{FBD7DA}62 & \cellcolor[HTML]{FBCCCE}73 & \cellcolor[HTML]{FBD1D4}68 & \cellcolor[HTML]{FBD1D3}68 & \cellcolor[HTML]{FBC2C5}81 & \cellcolor[HTML]{FBCBCE}73 & \cellcolor[HTML]{FBD3D5}66 & \cellcolor[HTML]{F98D8F}130 & \cellcolor[HTML]{FA9FA1}113 & \cellcolor[HTML]{FAA2A4}111 & \cellcolor[HTML]{F9888A}135 & \cellcolor[HTML]{FA9496}123 & \cellcolor[HTML]{FA9B9E}117 & \cellcolor[HTML]{F9797B}148 & \cellcolor[HTML]{F9888A}135 & \cellcolor[HTML]{FAA1A4}111 \\ \cline{2-20} 
\multicolumn{1}{|c|}{\multirow{-3}{*}{\rotatebox[origin=c]{90}{\textbf{MAT.}}}} & \textbf{A. EFF} & \cellcolor[HTML]{CEE1F5}1.4 & \cellcolor[HTML]{B8D3F1}1.7 & \cellcolor[HTML]{B9D4F1}1.7 & \cellcolor[HTML]{C0D8F2}1.6 & \cellcolor[HTML]{B4D1F0}\framebox{1.8} & \cellcolor[HTML]{B8D3F1}1.7 & \cellcolor[HTML]{CCDFF5}1.4 & \cellcolor[HTML]{BAD4F1}1.7 & \cellcolor[HTML]{C5DBF3}1.5 & \cellcolor[HTML]{73A8E2}2.8 & \cellcolor[HTML]{5A99DD}3.1 & \cellcolor[HTML]{619DDF}3.0 & \cellcolor[HTML]{659FDF}3.0 & \cellcolor[HTML]{5394DC}\framebox{3.2} & \cellcolor[HTML]{66A0E0}3.0 & \cellcolor[HTML]{7DAEE4}2.6 & \cellcolor[HTML]{5B99DD}3.1 & \cellcolor[HTML]{80B0E5}2.6 \\ \hline
\multicolumn{1}{|c|}{} & \textbf{PERF} & \cellcolor[HTML]{FCFCFF}0.53 & \cellcolor[HTML]{F5F9F9}0.69 & \cellcolor[HTML]{F4F9F9}0.69 & \cellcolor[HTML]{F9FBFD}0.59 & \cellcolor[HTML]{F2F8F6}0.74 & \cellcolor[HTML]{F3F9F8}0.71 & \cellcolor[HTML]{F8FBFC}0.62 & \cellcolor[HTML]{F0F8F5}0.77 & \cellcolor[HTML]{F4F9F8}0.70 & \cellcolor[HTML]{F8FAFB}0.63 & \cellcolor[HTML]{EDF6F2}0.85 & \cellcolor[HTML]{EEF7F3}0.82 & \cellcolor[HTML]{F5FAF9}0.68 & \cellcolor[HTML]{EBF5F0}0.89 & \cellcolor[HTML]{EDF6F3}0.83 & \cellcolor[HTML]{F4F9F8}0.69 & \cellcolor[HTML]{E9F5EF}0.91 & \cellcolor[HTML]{EEF7F3}0.81 \\ \cline{2-20} 
\multicolumn{1}{|c|}{} & \textbf{E. EFF} & \cellcolor[HTML]{FCF7FA}33 & \cellcolor[HTML]{FCFCFF}28 & \cellcolor[HTML]{FCFCFF}29 & \cellcolor[HTML]{FCF6F9}34 & \cellcolor[HTML]{FCFCFF}29 & \cellcolor[HTML]{FCFAFD}30 & \cellcolor[HTML]{FCF3F6}37 & \cellcolor[HTML]{FCF9FC}31 & \cellcolor[HTML]{FCFCFF}29 & \cellcolor[HTML]{FCF1F3}39 & \cellcolor[HTML]{FCF5F8}35 & \cellcolor[HTML]{FCF6F9}34 & \cellcolor[HTML]{FCF1F4}39 & \cellcolor[HTML]{FCF5F8}35 & \cellcolor[HTML]{FCF5F8}35 & \cellcolor[HTML]{FCEEF1}42 & \cellcolor[HTML]{FCF3F6}37 & \cellcolor[HTML]{FCF7FA}33 \\ \cline{2-20} 
\multicolumn{1}{|c|}{\multirow{-3}{*}{\rotatebox[origin=c]{90}{\textbf{SVM}}}} & \textbf{A. EFF} & \cellcolor[HTML]{FCFEFF}0.7 & \cellcolor[HTML]{F7FAFE}0.8 & \cellcolor[HTML]{F7FAFE}0.8 & \cellcolor[HTML]{F9FCFE}0.7 & \cellcolor[HTML]{F8FBFE}\framebox{0.8} & \cellcolor[HTML]{F8FBFE}0.8 & \cellcolor[HTML]{F9FCFE}0.7 & \cellcolor[HTML]{FBFDFF}0.7 & \cellcolor[HTML]{F9FCFE}0.7 & \cellcolor[HTML]{F4F8FD}0.8 & \cellcolor[HTML]{EBF3FB}1.0 & \cellcolor[HTML]{EDF4FC}0.9 & \cellcolor[HTML]{F2F7FD}0.9 & \cellcolor[HTML]{EEF5FC}\framebox{0.9} & \cellcolor[HTML]{F0F6FC}0.9 & \cellcolor[HTML]{FAFCFE}0.7 & \cellcolor[HTML]{F2F7FD}0.9 & \cellcolor[HTML]{F8FBFE}0.8 \\ \hline \hline 
\multicolumn{1}{|c|}{} & \textbf{PERF} & \cellcolor[HTML]{FCFCFF}0.00 & \cellcolor[HTML]{FCDADD}0.24 & \cellcolor[HTML]{FBD7DA}0.26 & \cellcolor[HTML]{FCE4E7}0.16 & \cellcolor[HTML]{FBC2C4}0.40 & \cellcolor[HTML]{FBC8CB}0.35 & \cellcolor[HTML]{FCDBDE}0.23 & \cellcolor[HTML]{FBB6B9}0.48 & \cellcolor[HTML]{FBC8CB}0.35 & \cellcolor[HTML]{FBC2C5}0.40 & \cellcolor[HTML]{F98D8F}0.76 & \cellcolor[HTML]{FA9193}0.73 & \cellcolor[HTML]{FAADB0}0.54 & \cellcolor[HTML]{F97577}0.92 & \cellcolor[HTML]{F98789}0.80 & \cellcolor[HTML]{FAA2A4}0.62 & \cellcolor[HTML]{F8696B}1.00 & \cellcolor[HTML]{F98A8C}0.78 \\ \cline{2-20} 
\multicolumn{1}{|c|}{} & \textbf{E. EFF} & \cellcolor[HTML]{EAF5F0}0.13 & \cellcolor[HTML]{FCFCFF}0.01 & \cellcolor[HTML]{F8FBFB}0.04 & \cellcolor[HTML]{D4ECDD}0.27 & \cellcolor[HTML]{EBF5F0}0.12 & \cellcolor[HTML]{E5F3EB}0.16 & \cellcolor[HTML]{BBE2C7}0.43 & \cellcolor[HTML]{D9EEE1}0.24 & \cellcolor[HTML]{EAF5EF}0.13 & \cellcolor[HTML]{8CCF9F}0.73 & \cellcolor[HTML]{ABDBB9}0.54 & \cellcolor[HTML]{ADDCBB}0.52 & \cellcolor[HTML]{83CB97}0.79 & \cellcolor[HTML]{9ED6AE}0.62 & \cellcolor[HTML]{9ED6AE}0.62 & \cellcolor[HTML]{63BE7B}1.00 & \cellcolor[HTML]{89CE9C}0.76 & \cellcolor[HTML]{A8DAB6}0.56 \\ \cline{2-20} 
\multicolumn{1}{|c|}{\multirow{-3}{*}{\rotatebox[origin=c]{90}{\textbf{NAVG}}}} & \textbf{A. EFF} & \cellcolor[HTML]{EAF2FA}0.03 & \cellcolor[HTML]{D4E5F5}0.20 & \cellcolor[HTML]{CCE0F3}0.23 & \cellcolor[HTML]{CADFF3}0.24 & \cellcolor[HTML]{BAD6EF}\framebox{\textbf{0.30}} & \cellcolor[HTML]{BBD7EF}0.29 & \cellcolor[HTML]{EAF2FA}0.12 & \cellcolor[HTML]{C1DAF1}0.27 & \cellcolor[HTML]{E4EFF9}0.14 & \cellcolor[HTML]{589BD8}0.66 & \cellcolor[HTML]{1774C8}0.91 & \cellcolor[HTML]{217ACB}0.87 & \cellcolor[HTML]{3284CF}0.80 & \cellcolor[HTML]{0F6FC6}\framebox{\textbf{0.94}} & \cellcolor[HTML]{257CCC}0.86 & \cellcolor[HTML]{65A3DB}0.61 & \cellcolor[HTML]{257CCC}0.85 & \cellcolor[HTML]{65A3DB}0.61 \\ \hline
\end{tabular}
}
\end{table*}
% Please add the following required packages to your document preamble:
% \usepackage{multirow}
% \usepackage[table.xcdraw]{xcolor}
% If you use beamer only pass "xcolor=table" option. i.e. \documentclass[xcolor=table]{beamer}
\begin{table*}[t]
\caption{Performance [$Gflop/s$], energy efficiency [$Gflop/s/W$], and area efficiency [$Gflop/s/mm^{2}$] executing the benchmarks on the 16-cores configurations. Performance and area efficiency are computed at 0.8 $V$, energy efficiency at 0.65 $V$. The last group of lines reports normalized average values.
A box around the metric value highlights the best configuration for each benchmark.}
\label{tab:results_16_cores}
\resizebox{\textwidth}{!}{
\begin{tabular}{cc|c|c|c|c|c|c|c|c|c|c|c|c|c|c|c|c|c|c|}
\cline{3-20}
 & \multicolumn{1}{c|}{} & \multicolumn{9}{c|}{\textbf{Scalar}} & \multicolumn{9}{c|}{\textbf{Vector}} \\ \cline{3-20} 
\multicolumn{2}{c}{} & \multicolumn{1}{|c|}{\textbf{16c4f0p}} & \multicolumn{1}{c|}{\textbf{16c4f1p}} & \multicolumn{1}{c|}{\textbf{16c4f2p}} & \multicolumn{1}{c|}{\textbf{16c8f0p}} & \multicolumn{1}{c|}{\textbf{16c8f1p}} & \multicolumn{1}{c|}{\textbf{16c8f2p}} & \multicolumn{1}{c|}{\textbf{16c16f0p}} & \multicolumn{1}{c|}{\textbf{16c16f1p}} & \multicolumn{1}{c|}{\textbf{16c16f2p}} & \multicolumn{1}{c|}{\textbf{16c4f0p}} & \multicolumn{1}{c|}{\textbf{16c4f1p}} & \multicolumn{1}{c|}{\textbf{16c4f2p}} & \multicolumn{1}{c|}{\textbf{16c8f0p}} & \multicolumn{1}{c|}{\textbf{16c8f1p}} & \multicolumn{1}{c|}{\textbf{16c8f2p}} & \multicolumn{1}{c|}{\textbf{16c16f0p}} & \multicolumn{1}{c|}{\textbf{16c16f1p}} & \multicolumn{1}{c|}{\textbf{16c16f2p}} \\ \hline
\multicolumn{1}{|c|}{} & \textbf{PERF} & \cellcolor[HTML]{D2EBDB}2.19 & \cellcolor[HTML]{C0E4CB}2.80 & \cellcolor[HTML]{BCE2C8}2.93 & \cellcolor[HTML]{C5E6D0}2.61 & \cellcolor[HTML]{B7E0C3}3.10 & \cellcolor[HTML]{B6E0C2}3.14 & \cellcolor[HTML]{C2E5CD}2.71 & \cellcolor[HTML]{AFDDBC}\framebox{3.37} & \cellcolor[HTML]{B2DEBF}3.26 & \cellcolor[HTML]{ABDBB9}3.51 & \cellcolor[HTML]{94D2A5}4.28 & \cellcolor[HTML]{A7DAB6}3.63 & \cellcolor[HTML]{A5D9B4}3.69 & \cellcolor[HTML]{8CCF9F}4.54 & \cellcolor[HTML]{A4D9B3}3.74 & \cellcolor[HTML]{9CD6AC}4.00 & \cellcolor[HTML]{85CC99}\framebox{4.78} & \cellcolor[HTML]{A0D7B0}3.87 \\ \cline{2-20} 
\multicolumn{1}{|c|}{} & \textbf{E. EFF} & \cellcolor[HTML]{FBC5C8}77 & \cellcolor[HTML]{FBCDD0}69 & \cellcolor[HTML]{FBCACD}72 & \cellcolor[HTML]{FBBEC0}84 & \cellcolor[HTML]{FBC7C9}75 & \cellcolor[HTML]{FBC9CC}73 & \cellcolor[HTML]{FBB3B6}\framebox{94} & \cellcolor[HTML]{FBC2C5}79 & \cellcolor[HTML]{FBC4C7}78 & \cellcolor[HTML]{FA9699}123 & \cellcolor[HTML]{FAA8AA}106 & \cellcolor[HTML]{FBB8BB}89 & \cellcolor[HTML]{FA9B9D}118 & \cellcolor[HTML]{FAA3A6}110 & \cellcolor[HTML]{FBBBBD}87 & \cellcolor[HTML]{F98588}\framebox{140} & \cellcolor[HTML]{FAA1A3}113 & \cellcolor[HTML]{FBB5B8}92 \\ \cline{2-20} 
\multicolumn{1}{|c|}{\multirow{-3}{*}{\rotatebox[origin=c]{90}{\textbf{CONV}}}} & \textbf{A. EFF} & \cellcolor[HTML]{C2E0FB}1.5 & \cellcolor[HTML]{B4D9FA}1.8 & \cellcolor[HTML]{A6D2F8}2.0 & \cellcolor[HTML]{B5D9FA}1.7 & \cellcolor[HTML]{B1D7F9}1.8 & \cellcolor[HTML]{AED6F9}1.9 & \cellcolor[HTML]{B9DCFA}1.7 & \cellcolor[HTML]{B3D8F9}1.8 & \cellcolor[HTML]{B0D7F9}1.8 & \cellcolor[HTML]{8BC4F6}2.5 & \cellcolor[HTML]{7CBCF5}2.7 & \cellcolor[HTML]{89C3F6}2.5 & \cellcolor[HTML]{8AC3F6}2.5 & \cellcolor[HTML]{7FBEF5}2.7 & \cellcolor[HTML]{98CBF7}2.2 & \cellcolor[HTML]{89C3F6}2.5 & \cellcolor[HTML]{86C1F6}2.5 & \cellcolor[HTML]{9CCCF8}2.2 \\ \hline
\multicolumn{1}{|c|}{} & \textbf{PERF} & \cellcolor[HTML]{FCFCFF}0.74 & \cellcolor[HTML]{F7FAFB}0.91 & \cellcolor[HTML]{F6FAFA}0.95 & \cellcolor[HTML]{F9FBFC}0.87 & \cellcolor[HTML]{F6FAFA}0.98 & \cellcolor[HTML]{F6FAFA}0.97 & \cellcolor[HTML]{F8FBFC}0.89 & \cellcolor[HTML]{F3F9F7}\framebox{1.06} & \cellcolor[HTML]{F5F9F9}1.00 & \cellcolor[HTML]{F8FBFC}0.88 & \cellcolor[HTML]{F3F9F7}1.07 & \cellcolor[HTML]{F2F8F6}1.10 & \cellcolor[HTML]{F7FAFB}0.94 & \cellcolor[HTML]{F3F9F8}1.05 & \cellcolor[HTML]{F3F9F8}1.06 & \cellcolor[HTML]{F7FAFB}0.94 & \cellcolor[HTML]{F2F8F6}\framebox{1.11} & \cellcolor[HTML]{F3F8F7}1.08 \\ \cline{2-20} 
\multicolumn{1}{|c|}{} & \textbf{E. EFF} & \cellcolor[HTML]{FCF9FC}26 & \cellcolor[HTML]{FCFCFF}22 & \cellcolor[HTML]{FCFBFE}23 & \cellcolor[HTML]{FCF7FA}28 & \cellcolor[HTML]{FCFBFE}24 & \cellcolor[HTML]{FCFCFF}23 & \cellcolor[HTML]{FCF4F7}\framebox{31} & \cellcolor[HTML]{FCFAFD}25 & \cellcolor[HTML]{FCFBFE}24 & \cellcolor[HTML]{FCF4F7}31 & \cellcolor[HTML]{FCF8FB}26 & \cellcolor[HTML]{FCF7FA}27 & \cellcolor[HTML]{FCF5F7}30 & \cellcolor[HTML]{FCF9FC}26 & \cellcolor[HTML]{FCFAFD}25 & \cellcolor[HTML]{FCF2F5}\framebox{33} & \cellcolor[HTML]{FCF8FB}26 & \cellcolor[HTML]{FCF9FC}26 \\ \cline{2-20} 
\multicolumn{1}{|c|}{\multirow{-3}{*}{\rotatebox[origin=c]{90}{\textbf{DWT}}}} & \textbf{A. EFF} & \cellcolor[HTML]{FEFFFF}0.5 & \cellcolor[HTML]{FBFDFF}0.6 & \cellcolor[HTML]{F7FBFF}0.6 & \cellcolor[HTML]{FBFDFF}0.6 & \cellcolor[HTML]{FBFDFF}0.6 & \cellcolor[HTML]{FBFDFF}0.6 & \cellcolor[HTML]{FDFEFF}0.5 & \cellcolor[HTML]{FCFEFF}0.6 & \cellcolor[HTML]{FCFEFF}0.6 & \cellcolor[HTML]{F9FCFF}0.6 & \cellcolor[HTML]{F5FAFF}0.7 & \cellcolor[HTML]{F0F8FE}0.8 & \cellcolor[HTML]{F8FCFF}0.6 & \cellcolor[HTML]{F9FCFF}0.6 & \cellcolor[HTML]{F8FCFF}0.6 & \cellcolor[HTML]{FBFDFF}0.6 & \cellcolor[HTML]{FAFDFF}0.6 & \cellcolor[HTML]{F9FCFF}0.6 \\ \hline
\multicolumn{1}{|c|}{} & \textbf{PERF} & \cellcolor[HTML]{EFF7F4}1.21 & \cellcolor[HTML]{E6F3EC}1.51 & \cellcolor[HTML]{E4F3EB}1.56 & \cellcolor[HTML]{E6F3EC}1.52 & \cellcolor[HTML]{DEF0E5}1.78 & \cellcolor[HTML]{DDF0E4}1.81 & \cellcolor[HTML]{E3F2EA}1.60 & \cellcolor[HTML]{D8EEE0}\framebox{1.99} & \cellcolor[HTML]{DAEFE2}1.90 & \cellcolor[HTML]{D4ECDC}2.13 & \cellcolor[HTML]{C8E7D2}2.54 & \cellcolor[HTML]{C9E7D3}2.50 & \cellcolor[HTML]{D0EAD9}2.25 & \cellcolor[HTML]{C5E6D0}2.62 & \cellcolor[HTML]{C8E7D2}2.53 & \cellcolor[HTML]{D1EBDA}2.22 & \cellcolor[HTML]{C2E5CD}\framebox{2.74} & \cellcolor[HTML]{C6E6D1}2.58 \\ \cline{2-20} 
\multicolumn{1}{|c|}{} & \textbf{E. EFF} & \cellcolor[HTML]{FCE8EB}42 & \cellcolor[HTML]{FCEDF0}37 & \cellcolor[HTML]{FCECEF}38 & \cellcolor[HTML]{FCE2E4}49 & \cellcolor[HTML]{FCE7EA}43 & \cellcolor[HTML]{FCE8EB}42 & \cellcolor[HTML]{FCDADD}\framebox{56} & \cellcolor[HTML]{FCE4E6}47 & \cellcolor[HTML]{FCE5E8}45 & \cellcolor[HTML]{FBC7CA}74 & \cellcolor[HTML]{FBD4D6}62 & \cellcolor[HTML]{FBD5D7}61 & \cellcolor[HTML]{FBCACC}72 & \cellcolor[HTML]{FBD3D5}63 & \cellcolor[HTML]{FBD7DA}59 & \cellcolor[HTML]{FBC4C7}\framebox{78} & \cellcolor[HTML]{FBD2D4}64 & \cellcolor[HTML]{FBD5D7}62 \\ \cline{2-20} 
\multicolumn{1}{|c|}{\multirow{-3}{*}{\rotatebox[origin=c]{90}{\textbf{FFT}}}} & \textbf{A. EFF} & \cellcolor[HTML]{EBF5FE}0.8 & \cellcolor[HTML]{E4F2FD}1.0 & \cellcolor[HTML]{DEEEFD}1.1 & \cellcolor[HTML]{E1F0FD}1.0 & \cellcolor[HTML]{DFEFFD}1.0 & \cellcolor[HTML]{DDEEFD}1.1 & \cellcolor[HTML]{E2F1FD}1.0 & \cellcolor[HTML]{DFEFFD}1.1 & \cellcolor[HTML]{DEEEFD}1.1 & \cellcolor[HTML]{C5E1FB}1.5 & \cellcolor[HTML]{BEDEFA}1.6 & \cellcolor[HTML]{B7DBFA}1.7 & \cellcolor[HTML]{C4E1FB}1.5 & \cellcolor[HTML]{C2E0FB}1.5 & \cellcolor[HTML]{C3E1FB}1.5 & \cellcolor[HTML]{CBE5FB}1.4 & \cellcolor[HTML]{C7E2FB}1.5 & \cellcolor[HTML]{C7E2FB}1.5 \\ \hline
\multicolumn{1}{|c|}{} & \textbf{PERF} & \cellcolor[HTML]{CFEAD8}2.29 & \cellcolor[HTML]{C4E6CF}2.66 & \cellcolor[HTML]{CCE9D6}2.38 & \cellcolor[HTML]{C2E5CD}2.71 & \cellcolor[HTML]{BEE3CA}2.86 & \cellcolor[HTML]{CCE9D5}2.39 & \cellcolor[HTML]{BEE3CA}2.85 & \cellcolor[HTML]{B7E1C4}\framebox{3.08} & \cellcolor[HTML]{CAE8D4}2.47 & \cellcolor[HTML]{97D4A8}4.17 & \cellcolor[HTML]{79C78E}5.19 & \cellcolor[HTML]{8ACE9D}4.62 & \cellcolor[HTML]{8ACE9D}4.62 & \cellcolor[HTML]{73C589}5.38 & \cellcolor[HTML]{8CCF9F}4.54 & \cellcolor[HTML]{85CC98}4.79 & \cellcolor[HTML]{63BE7B}\framebox{5.92} & \cellcolor[HTML]{8ACE9D}4.62 \\ \cline{2-20} 
\multicolumn{1}{|c|}{} & \textbf{E. EFF} & \cellcolor[HTML]{FBC2C4}80 & \cellcolor[HTML]{FBD0D3}66 & \cellcolor[HTML]{FBD8DA}59 & \cellcolor[HTML]{FBBBBD}87 & \cellcolor[HTML]{FBCDCF}69 & \cellcolor[HTML]{FCDADD}56 & \cellcolor[HTML]{FAAEB1}\framebox{99} & \cellcolor[HTML]{FBC9CC}73 & \cellcolor[HTML]{FBD7DA}59 & \cellcolor[HTML]{F97F81}146 & \cellcolor[HTML]{FA9193}128 & \cellcolor[HTML]{FAA0A2}114 & \cellcolor[HTML]{F97D7F}148 & \cellcolor[HTML]{FA8F91}130 & \cellcolor[HTML]{FAA8AA}106 & \cellcolor[HTML]{F8696B}\framebox{167} & \cellcolor[HTML]{F98588}139 & \cellcolor[HTML]{FAA3A6}110 \\ \cline{2-20} 
\multicolumn{1}{|c|}{\multirow{-3}{*}{\rotatebox[origin=c]{90}{\textbf{FIR}}}} & \textbf{A. EFF} & \cellcolor[HTML]{BEDEFA}1.6 & \cellcolor[HTML]{B9DBFA}1.7 & \cellcolor[HTML]{BCDDFA}1.6 & \cellcolor[HTML]{B1D7F9}1.8 & \cellcolor[HTML]{BADCFA}1.7 & \cellcolor[HTML]{C8E3FB}1.4 & \cellcolor[HTML]{B4D9FA}1.8 & \cellcolor[HTML]{BCDDFA}1.6 & \cellcolor[HTML]{CBE4FB}1.4 & \cellcolor[HTML]{70B6F4}2.9 & \cellcolor[HTML]{59AAF2}3.3 & \cellcolor[HTML]{61AEF3}3.2 & \cellcolor[HTML]{65B0F3}3.1 & \cellcolor[HTML]{62AFF3}3.1 & \cellcolor[HTML]{7CBCF5}2.7 & \cellcolor[HTML]{6CB4F4}3.0 & \cellcolor[HTML]{62AFF3}3.1 & \cellcolor[HTML]{82BFF6}2.6 \\ \hline
\multicolumn{1}{|c|}{} & \textbf{PERF} & \cellcolor[HTML]{FBFCFF}0.78 & \cellcolor[HTML]{F6FAFA}0.95 & \cellcolor[HTML]{F5FAF9}\framebox{0.99} & \cellcolor[HTML]{FAFCFE}0.81 & \cellcolor[HTML]{F7FAFB}0.93 & \cellcolor[HTML]{F6FAFA}0.95 & \cellcolor[HTML]{FBFCFE}0.81 & \cellcolor[HTML]{F5FAF9}\framebox{0.98} & \cellcolor[HTML]{F6FAFA}0.97 & \cellcolor[HTML]{EAF5F0}1.37 & \cellcolor[HTML]{E1F1E7}1.69 & \cellcolor[HTML]{DFF1E6}\framebox{1.73} & \cellcolor[HTML]{E8F4EE}1.42 & \cellcolor[HTML]{E2F2E9}1.62 & \cellcolor[HTML]{E2F2E8}1.65 & \cellcolor[HTML]{E9F5EF}1.41 & \cellcolor[HTML]{E0F1E7}\framebox{1.71} & \cellcolor[HTML]{E1F1E7}1.68 \\ \cline{2-20} 
\multicolumn{1}{|c|}{} & \textbf{E. EFF} & \cellcolor[HTML]{FCF7FA}27 & \cellcolor[HTML]{FCFBFE}23 & \cellcolor[HTML]{FCFAFD}24 & \cellcolor[HTML]{FCF9FB}26 & \cellcolor[HTML]{FCFCFF}23 & \cellcolor[HTML]{FCFCFF}22 & \cellcolor[HTML]{FCF6F9}\framebox{28} & \cellcolor[HTML]{FCFCFF}23 & \cellcolor[HTML]{FCFBFE}23 & \cellcolor[HTML]{FCE2E5}48 & \cellcolor[HTML]{FCE9EC}42 & \cellcolor[HTML]{FCE8EB}42 & \cellcolor[HTML]{FCE5E8}46 & \cellcolor[HTML]{FCEBEE}39 & \cellcolor[HTML]{FCECEF}39 & \cellcolor[HTML]{FCE1E4}\framebox{49} & \cellcolor[HTML]{FCEAED}40 & \cellcolor[HTML]{FCEAED}40 \\ \cline{2-20} 
\multicolumn{1}{|c|}{\multirow{-3}{*}{\rotatebox[origin=c]{90}{\textbf{IIR}}}} & \textbf{A. EFF} & \cellcolor[HTML]{FDFEFF}0.5 & \cellcolor[HTML]{FAFCFF}0.6 & \cellcolor[HTML]{F5FAFF}0.7 & \cellcolor[HTML]{FDFEFF}0.5 & \cellcolor[HTML]{FDFEFF}0.5 & \cellcolor[HTML]{FBFDFF}0.6 & \cellcolor[HTML]{FDFEFF}0.5 & \cellcolor[HTML]{FEFFFF}0.5 & \cellcolor[HTML]{FDFEFF}0.5 & \cellcolor[HTML]{E4F2FD}1.0 & \cellcolor[HTML]{DEEEFD}1.1 & \cellcolor[HTML]{D7EBFC}1.2 & \cellcolor[HTML]{E5F2FD}1.0 & \cellcolor[HTML]{E5F2FD}0.9 & \cellcolor[HTML]{E2F1FD}1.0 & \cellcolor[HTML]{E9F4FE}0.9 & \cellcolor[HTML]{E7F3FE}0.9 & \cellcolor[HTML]{E5F2FD}0.9 \\ \hline
\multicolumn{1}{|c|}{} & \textbf{PERF} & \cellcolor[HTML]{F1F8F5}1.14 & \cellcolor[HTML]{E9F5EF}1.39 & \cellcolor[HTML]{E9F5EF}1.39 & \cellcolor[HTML]{EDF6F2}1.28 & \cellcolor[HTML]{E8F4EE}1.45 & \cellcolor[HTML]{EAF5F0}1.35 & \cellcolor[HTML]{EEF6F3}1.25 & \cellcolor[HTML]{E6F3EC}\framebox{1.50} & \cellcolor[HTML]{E9F5EF}1.40 & \cellcolor[HTML]{E0F1E7}1.72 & \cellcolor[HTML]{D4ECDD}2.11 & \cellcolor[HTML]{D3ECDC}2.13 & \cellcolor[HTML]{D9EEE1}1.95 & \cellcolor[HTML]{CFEAD8}2.28 & \cellcolor[HTML]{D1EBDA}2.22 & \cellcolor[HTML]{D9EEE1}1.93 & \cellcolor[HTML]{CBE8D5}\framebox{2.43} & \cellcolor[HTML]{CFEAD8}2.29 \\ \cline{2-20} 
\multicolumn{1}{|c|}{} & \textbf{E. EFF} & \cellcolor[HTML]{FCEAED}40 & \cellcolor[HTML]{FCF0F3}34 & \cellcolor[HTML]{FCF0F3}34 & \cellcolor[HTML]{FCE9EC}41 & \cellcolor[HTML]{FCEFF2}35 & \cellcolor[HTML]{FCF3F6}32 & \cellcolor[HTML]{FCE7EA}\framebox{43} & \cellcolor[HTML]{FCEFF2}35 & \cellcolor[HTML]{FCF1F4}33 & \cellcolor[HTML]{FBD6D9}60 & \cellcolor[HTML]{FCDEE1}52 & \cellcolor[HTML]{FCDEE1}52 & \cellcolor[HTML]{FBD4D6}63 & \cellcolor[HTML]{FCDBDE}55 & \cellcolor[HTML]{FCDEE1}52 & \cellcolor[HTML]{FBCFD1}\framebox{67} & \cellcolor[HTML]{FCD9DC}57 & \cellcolor[HTML]{FCDCDF}54 \\ \cline{2-20} 
\multicolumn{1}{|c|}{\multirow{-3}{*}{\rotatebox[origin=c]{90}{\textbf{K-M.}}}} & \textbf{A. EFF} & \cellcolor[HTML]{EEF6FE}0.8 & \cellcolor[HTML]{E9F4FE}0.9 & \cellcolor[HTML]{E4F2FD}1.0 & \cellcolor[HTML]{EAF5FE}0.9 & \cellcolor[HTML]{EBF5FE}0.8 & \cellcolor[HTML]{EDF6FE}0.8 & \cellcolor[HTML]{EFF7FE}0.8 & \cellcolor[HTML]{EEF7FE}0.8 & \cellcolor[HTML]{EFF7FE}0.8 & \cellcolor[HTML]{D6EAFC}1.2 & \cellcolor[HTML]{CEE6FC}1.3 & \cellcolor[HTML]{C6E2FB}1.5 & \cellcolor[HTML]{D0E7FC}1.3 & \cellcolor[HTML]{CEE6FC}1.3 & \cellcolor[HTML]{CEE6FC}1.3 & \cellcolor[HTML]{D6EAFC}1.2 & \cellcolor[HTML]{D1E7FC}1.3 & \cellcolor[HTML]{D1E8FC}1.3 \\ \hline
\multicolumn{1}{|c|}{} & \textbf{PERF} & \cellcolor[HTML]{D9EEE1}1.96 & \cellcolor[HTML]{C7E7D1}2.57 & \cellcolor[HTML]{CBE9D5}2.41 & \cellcolor[HTML]{D0EBDA}2.23 & \cellcolor[HTML]{C4E6CF}2.65 & \cellcolor[HTML]{CBE9D5}2.41 & \cellcolor[HTML]{CEEAD8}2.30 & \cellcolor[HTML]{BEE3CA}\framebox{2.86} & \cellcolor[HTML]{C8E7D3}2.50 & \cellcolor[HTML]{9DD6AD}3.98 & \cellcolor[HTML]{84CC97}4.83 & \cellcolor[HTML]{8BCF9E}4.57 & \cellcolor[HTML]{92D1A4}4.34 & \cellcolor[HTML]{7BC88F}5.14 & \cellcolor[HTML]{8ED0A0}4.48 & \cellcolor[HTML]{90D0A2}4.42 & \cellcolor[HTML]{71C487}\framebox{5.47} & \cellcolor[HTML]{8CCF9E}4.57 \\ \cline{2-20} 
\multicolumn{1}{|c|}{} & \textbf{E. EFF} & \cellcolor[HTML]{FBCED0}68 & \cellcolor[HTML]{FBD3D6}63 & \cellcolor[HTML]{FBD7D9}59 & \cellcolor[HTML]{FBCACD}72 & \cellcolor[HTML]{FBD2D5}64 & \cellcolor[HTML]{FCDADD}56 & \cellcolor[HTML]{FBC2C4}\framebox{80} & \cellcolor[HTML]{FBCFD1}67 & \cellcolor[HTML]{FBD6D9}60 & \cellcolor[HTML]{F98688}139 & \cellcolor[HTML]{FA9A9D}119 & \cellcolor[HTML]{FAA1A3}113 & \cellcolor[HTML]{F98588}139 & \cellcolor[HTML]{FA9597}125 & \cellcolor[HTML]{FAA9AC}104 & \cellcolor[HTML]{F97678}\framebox{154} & \cellcolor[HTML]{FA9093}129 & \cellcolor[HTML]{FAA5A7}109 \\ \cline{2-20} 
\multicolumn{1}{|c|}{\multirow{-3}{*}{\rotatebox[origin=c]{90}{\textbf{MAT.}}}} & \textbf{A. EFF} & \cellcolor[HTML]{CCE5FB}1.4 & \cellcolor[HTML]{BDDDFA}1.6 & \cellcolor[HTML]{BBDCFA}1.6 & \cellcolor[HTML]{C4E1FB}1.5 & \cellcolor[HTML]{C1DFFB}1.5 & \cellcolor[HTML]{C8E3FB}1.4 & \cellcolor[HTML]{C8E3FB}1.4 & \cellcolor[HTML]{C3E0FB}1.5 & \cellcolor[HTML]{C9E4FB}1.4 & \cellcolor[HTML]{77BAF5}2.8 & \cellcolor[HTML]{67B2F4}3.0 & \cellcolor[HTML]{62AFF3}3.1 & \cellcolor[HTML]{70B6F4}2.9 & \cellcolor[HTML]{6AB3F4}3.0 & \cellcolor[HTML]{7EBDF5}2.7 & \cellcolor[HTML]{7ABBF5}2.7 & \cellcolor[HTML]{70B6F4}2.9 & \cellcolor[HTML]{84C0F6}2.6 \\ \hline
\multicolumn{1}{|c|}{} & \textbf{PERF} & \cellcolor[HTML]{F8FBFC}0.90 & \cellcolor[HTML]{F3F9F7}1.07 & \cellcolor[HTML]{F3F9F7}1.06 & \cellcolor[HTML]{F5FAF9}0.99 & \cellcolor[HTML]{F2F8F6}1.11 & \cellcolor[HTML]{F3F9F7}1.07 & \cellcolor[HTML]{F5F9F9}1.00 & \cellcolor[HTML]{EFF7F4}\framebox{1.19} & \cellcolor[HTML]{F2F8F7}1.09 & \cellcolor[HTML]{F1F8F5}1.14 & \cellcolor[HTML]{E9F5EF}1.40 & \cellcolor[HTML]{E9F5EF}1.39 & \cellcolor[HTML]{EFF7F3}1.21 & \cellcolor[HTML]{EAF5EF}1.38 & \cellcolor[HTML]{EBF5F0}1.33 & \cellcolor[HTML]{EFF7F4}1.21 & \cellcolor[HTML]{E7F4ED}\framebox{1.47} & \cellcolor[HTML]{EAF5F0}1.36 \\ \cline{2-20} 
\multicolumn{1}{|c|}{} & \textbf{E. EFF} & \cellcolor[HTML]{FCF3F6}31 & \cellcolor[HTML]{FCF8FB}26 & \cellcolor[HTML]{FCF8FB}26 & \cellcolor[HTML]{FCF3F6}32 & \cellcolor[HTML]{FCF8FB}27 & \cellcolor[HTML]{FCFAFD}25 & \cellcolor[HTML]{FCF0F3}\framebox{35} & \cellcolor[HTML]{FCF7FA}28 & \cellcolor[HTML]{FCF9FC}26 & \cellcolor[HTML]{FCEAED}40 & \cellcolor[HTML]{FCF0F3}34 & \cellcolor[HTML]{FCF0F3}34 & \cellcolor[HTML]{FCEBEE}39 & \cellcolor[HTML]{FCF1F4}34 & \cellcolor[HTML]{FCF4F6}31 & \cellcolor[HTML]{FCE8EB}\framebox{42} & \cellcolor[HTML]{FCF0F3}35 & \cellcolor[HTML]{FCF2F5}32 \\ \cline{2-20} 
\multicolumn{1}{|c|}{\multirow{-3}{*}{\rotatebox[origin=c]{90}{\textbf{SVM}}}} & \textbf{A. EFF} & \cellcolor[HTML]{F8FCFF}0.6 & \cellcolor[HTML]{F5FAFF}0.7 & \cellcolor[HTML]{F2F9FE}0.7 & \cellcolor[HTML]{F6FBFF}0.7 & \cellcolor[HTML]{F7FBFF}0.6 & \cellcolor[HTML]{F7FBFF}0.6 & \cellcolor[HTML]{F8FCFF}0.6 & \cellcolor[HTML]{F8FCFF}0.6 & \cellcolor[HTML]{F9FCFF}0.6 & \cellcolor[HTML]{EEF6FE}0.8 & \cellcolor[HTML]{E9F4FE}0.9 & \cellcolor[HTML]{E5F2FD}1.0 & \cellcolor[HTML]{EDF6FE}0.8 & \cellcolor[HTML]{EDF6FE}0.8 & \cellcolor[HTML]{EEF7FE}0.8 & \cellcolor[HTML]{F1F8FE}0.7 & \cellcolor[HTML]{EFF7FE}0.8 & \cellcolor[HTML]{F0F8FE}0.8 \\ \hline \hline 
\multicolumn{1}{|c|}{} & \textbf{PERF} & \cellcolor[HTML]{FCFCFF}0.00 & \cellcolor[HTML]{D9EEE1}0.23 & \cellcolor[HTML]{D8EEE0}0.24 & \cellcolor[HTML]{E6F3EC}0.15 & \cellcolor[HTML]{CDE9D7}0.31 & \cellcolor[HTML]{D3ECDC}0.27 & \cellcolor[HTML]{E2F2E9}0.17 & \cellcolor[HTML]{BDE3C9}\framebox{\textbf{0.41}} & \cellcolor[HTML]{CDE9D6}0.31 & \cellcolor[HTML]{AEDDBC}0.51 & \cellcolor[HTML]{7BC890}0.84 & \cellcolor[HTML]{82CB96}0.80 & \cellcolor[HTML]{9DD6AE}0.62 & \cellcolor[HTML]{75C68B}0.88 & \cellcolor[HTML]{86CD99}0.77 & \cellcolor[HTML]{9BD5AB}0.64 & \cellcolor[HTML]{63BE7B}\framebox{\textbf{1.00}} & \cellcolor[HTML]{80CA94}0.81 \\ \cline{2-20} 
\multicolumn{1}{|c|}{} & \textbf{E. EFF} & \cellcolor[HTML]{FCDFE2}0.22 & \cellcolor[HTML]{FCF9FC}0.05 & \cellcolor[HTML]{FCF8FB}0.06 & \cellcolor[HTML]{FBD5D8}0.28 & \cellcolor[HTML]{FCF1F4}0.10 & \cellcolor[HTML]{FCFCFF}0.02 & \cellcolor[HTML]{FBC0C2}\framebox{\textbf{0.43}} & \cellcolor[HTML]{FCE9EC}0.15 & \cellcolor[HTML]{FCF4F6}0.08 & \cellcolor[HTML]{F98082}0.85 & \cellcolor[HTML]{FAA8AB}0.58 & \cellcolor[HTML]{FAAFB1}0.54 & \cellcolor[HTML]{F98587}0.82 & \cellcolor[HTML]{FAA8AA}0.59 & \cellcolor[HTML]{FBBFC1}0.43 & \cellcolor[HTML]{F8696B}\framebox{\textbf{1.00}} & \cellcolor[HTML]{FAA0A2}0.64 & \cellcolor[HTML]{FBB4B7}0.50 \\ \cline{2-20} 
\multicolumn{1}{|c|}{\multirow{-3}{*}{\rotatebox[origin=c]{90}{\textbf{NAVG}}}} & \textbf{A. EFF} & \cellcolor[HTML]{F2F7FC}0.03 & \cellcolor[HTML]{DCEAF7}0.17 & \cellcolor[HTML]{BCD7F0}0.29 & \cellcolor[HTML]{DFECF8}0.16 & \cellcolor[HTML]{E1EDF8}0.15 & \cellcolor[HTML]{E4EFF9}0.14 & \cellcolor[HTML]{F2F7FC}0.08 & \cellcolor[HTML]{E9F2FA}0.12 & \cellcolor[HTML]{EEF5FB}0.10 & \cellcolor[HTML]{5B9DD8}0.67 & \cellcolor[HTML]{2A7FCD}0.87 & \cellcolor[HTML]{0F6FC6}0.97 & \cellcolor[HTML]{4D95D5}0.73 & \cellcolor[HTML]{4590D3}0.76 & \cellcolor[HTML]{5D9ED9}0.67 & \cellcolor[HTML]{6CA7DD}0.61 & \cellcolor[HTML]{589BD8}0.69 & \cellcolor[HTML]{6FA9DD}0.60 \\ \hline

\end{tabular}
}
\end{table*}

%%%%%%%%%%%%%%%%%% FIR/IIR %%%%%%%%%%%%%%%%%%
The Finite Impulse Response (FIR) and Infinite Impulse Response (IIR) are digital filters with various applications in data acquisition and analysis.
%The FIR has a finite impulse response, depending on the current input and a fixed number of previous data. Conversely, the IIR filter has an infinite impulse response depending on the current input, a fixed number of prior data, and a fixed number of previous outputs.
%
%%%%%%%%%%%%%%%%%% DWT %%%%%%%%%%%%%%%%%%
%\cite{montagna2017flexible}
The Discrete Wavelet Transform (DWT) is a standard kernel used for feature extraction, which decomposes a signal into a different level of frequency resolutions through a bank of Low Pass (LPF) and High Pass Filters (HPF), capturing both temporal and frequency information.
%%%%%%%%%%%%%%%%%% FFT %%%%%%%%%%%%%%%%%%
%
%\cite{kartsch2018sensor}
The Fast Fourier Transform (FFT) is a mathematical method that transforms a signal from the time domain to the frequency domain.
%This algorithm requires to compute a set of \textit{butterflies}, which are computations that combine the results of an algorithmic stage into the next one. The FFT requires to compute a variable set of stages, depending on the input size. 
There are several variants of this algorithm; in this paper, we consider the decimation-in-frequency radix-2 variant.
%
%%%%%%%%%%%%%%%%%% SVM %%%%%%%%%%%%%%%%%%
We consider a state-of-the-art supervised classifier, the Support Vector Machine (SVM), widely used in near-sensor applications \cite{montagna2017machine}. %Starting from a set of support vectors (SVs) that define a boundary hyper-plane, it classifies unknown samples into a known class.
%The SVM training defines a set of support vectors (SVs) defining a boundary hyper-plane that separates the training dataset in 2 classes. 
%The SVM inference maps an input sample on a higher-dimensional space, computing the kernel function and calculates the distance between the input sample and the hyper-plane. 
%
%%%%%%%%%%%%%%%%%% K-MEANS %%%%%%%%%%%%%%%%%%
We also include another classifier, named K-Means, which is an unsupervised ML algorithm able to inference an unknown outcome starting from input vectors.
%This is done grouping similar samples together into a given number of clusters (\textit{K}), discovering hidden patterns.
%
The last two kernels are basic linear algebra subprograms (BLAS) commonly used in DSP: matrix multiplication (MATMUL) and convolution (CONV), which is the most computing-intensive kernel in convolutional neural network (CNN) workloads.

% Scalar and vector variants
We have implemented different variants of each kernel, using scalar (\emph{float}) and vector (2 $\times$ \emph{float16}, 2 $\times$ \emph{bfloat16}) data types.
Considering the design of the FPU, there is no significant difference in execution time and energy consumption between \emph{float16} and \emph{bfloat16} vectors; in the following experiments, we report a single value for both configurations.

% Parallelism
To exploit the parallelism provided by the transprecision cluster, each variant accepts a parameter representing the number of cores available in the current configuration.
The source code includes a form of parametric parallelism based on the number of available cores and the core id, using the low-overhead HAL interface described in Section~\ref{sec:software}.

We exploited data parallelism at loop level with static scheduling of the iterations on the available cores.
This policy guarantees the maximum balancing with a limited overhead related to the computation of per-core iteration boundaries.
Whenever it is feasible, we apply data parallelism to the outer loops of the benchmarks (CONV, FIR, MATMUL). In other cases, data parallelism is applied to single stages of the algorithm, separated by a synchronization barrier (DWT, FFT, KMEANS, SVM); except for FFT, these benchmarks are characterized by sequential regions interleaved with parallel loops and executed by a single core.

A common problem of IIR filters working on a single stream is that data dependencies limit the parallelism.
To alleviate this limitation, we have adopted a technique based on a block formulation of recursive filters tailored for vector units \cite{robelly2004implementation}.
The algebraic transformations required by this technique are applied off-line and do not imply any overhead. However, the time complexity of the algorithm is higher than the original one and the size of the vector state (equal to the number of taps) severely limits the exploitability of parallelism.
For this reason, the vector variant of this benchmark is the only reported case with alternative configurations achieving the best result for energy efficiency.

% FP Intensity and Memory intensity
Table \ref{tab:tab_app} also reports the FP and memory intensity of the benchmarks for scalar and vector variants.
The FP intensity is computed as the ratio between the number of FP instructions and the total number of instructions.
Analogously, the memory intensity is the number of load/store instructions over the total number of instructions.
These numbers provide a quantitative evaluation of the pressure on the FPU and memory subsystems and are essential to understand the actual FP workload in a real execution scenario.
%
%\begin{figure*}[t]
%\begin{center}
%\includegraphics[width=1\textwidth]{FIGS/SPEED-UP-8-CORES.eps}
%%\vspace{-6mm}
%\caption{Speed-ups obtained by the DSP benchmarks, executing the f32 and f16 vectorial implementations on all the 8-cores configurations in the design space. Each bar shows the minimum (dark color), maximum and average (light color) values.}
%%\vspace{-3mm}
%\label{fig:SPEEDUP8CORES}
%\end{center}
%\end{figure*} 
%
%
\begin{figure*}[t]
\begin{center}
\includegraphics[width=\textwidth]{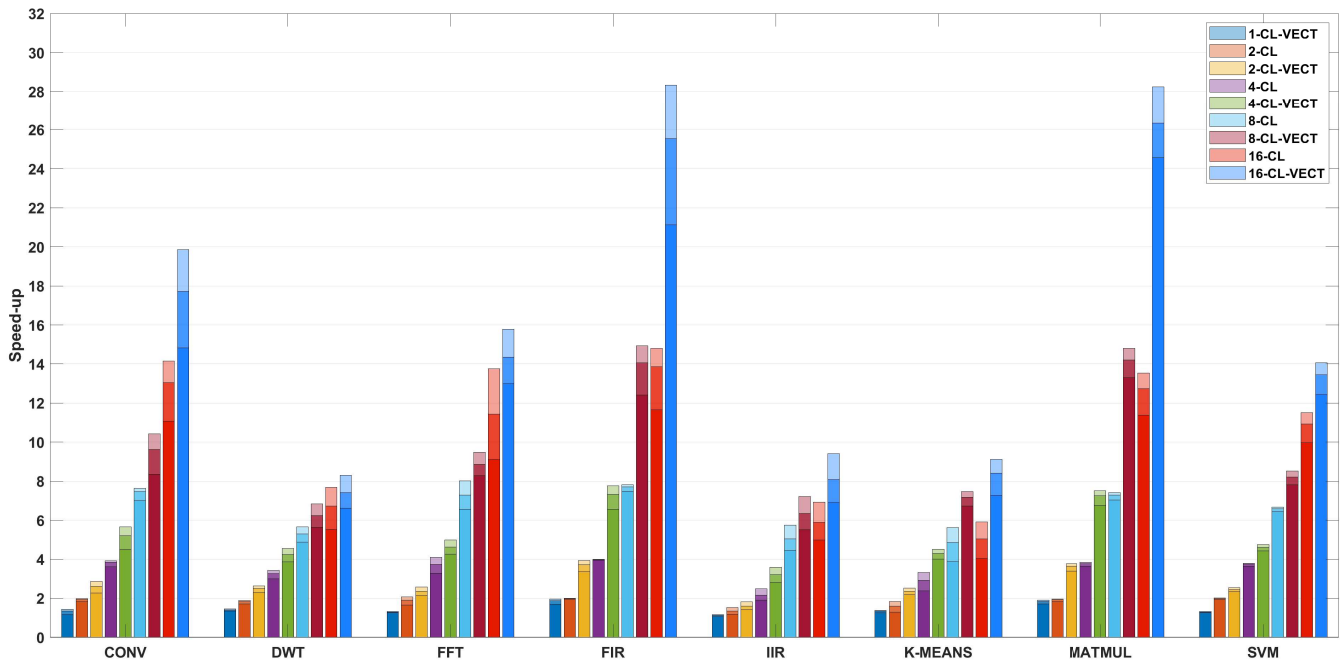}
%\vspace{-6mm}
\caption{Speed-ups obtained executing scalar and vector variants on all the platform configurations. Each configuration reports the number of available cores and the support to vectorization. Each bar shows the minimum (dark color), maximum and average (light color) value.}
%\vspace{-3mm}
\label{fig:SPEEDUP16CORES}
\end{center}
\end{figure*}
\subsection{Benchmarking}
\label{sec:benchmarking}
We have performed an extensive benchmarking considering all the benchmark variants and all the configurations of the transprecision cluster.
We have measured the performance (Gflop/s), the energy efficiency (Gflop/s/W), and the area efficiency (Gflop/s/mm$^2$) for each benchmark variant and platform configuration: Table~\ref{tab:results_8_cores} and Table~\ref{tab:results_16_cores} report the result of the experiments.
The last row of each table reports the normalized average of the measures, computed normalizing the values into values between 0 and 1 (min-max normalization).
Computing these metrics allows establishing the configurations that, on average, return the worst and the best performance and energy/area efficiency for the considered benchmarks.
Moreover, the tables use a color scale to visually emphasize the worst (light color) and the best (dark color) configurations. 

The configuration with 16 cores, private FPUs, and one pipeline stage provides the best performance, with a maximum of 3.37 Gflop/s for scalars and 5.92 Gflop/s for vectors. It is quite intuitive that using the maximum number of cores and FPUs is beneficial for performance. An additional pipeline stage could enable a further increase of the frequency, but this is not the case due to structural critical paths (as discussed in Section~\ref{sec:cluster_impl}.

The configuration with 16 cores, private FPUs, and zero pipeline stages is the most energy-efficient, with a maximum of 99 Gflop/s/W and 167 Gflop/s/W for vectors. Using the maximum number of cores is never detrimental to performance, mainly thanks to the adoption of aggressive power-saving policies that turn off cores waiting for synchronization events.
Moreover, this configuration removes FPU stalls, which are detrimental to energy efficiency.

The configuration with 8 cores and 4 shared FPUs configured with one pipeline stage is the most area-efficient, with a maximum of 2.0 Gflop/s/mm$^2$ for scalars and 3.5 Gflop/s/mm$^2$ for vectors.
This configuration saves area reducing the number of cores and the sharing factor, but maintaining a single pipeline stage represents the best tradeoff with performance.
%
%\begin{figure}[t]
%\begin{center}
%\includegraphics[width=0.99\columnwidth]{FIGS/prova_cyc.png}
%\vspace{-6mm}
%\caption{Number of cycles for the DSP benchmarks normalized to the worst case - float32. The results refer to the 8-cores, 8 FPUs and 0 pipeline stages.}
%\vspace{-3mm}
%\label{fig:32_16_vect}
%\end{center}
%\end{figure} 
\subsubsection{Parallelization and vectorization}
Fig. \ref{fig:SPEEDUP16CORES} depicts the speed-ups from the execution of the benchmarks on the 16-cores architectures, combining the benefits deriving from parallelism and vectorization.
Each configuration of the transprecision cluster is denoted by the abbreviation $n$-CL, where $n$ indicates the number of cores.
The suffix VECT designates the execution of the vector variant.
The baseline to compute the speed-up is the execution on a single core with no vectorization support.
The bars show average, maximum, and minimum values of the speed-ups executed on all the architectural configurations.

Focusing on the parallel speed-up, we can notice that the values reported for DWT, IIR, and K-MEANS are modest, reaching a saturation point around 8.
These benchmarks have a complex parallel execution flow, requiring several synchronization barriers and regions with sequential execution to ensure the correctness of the results, and this limits the parallelism.
However, this effect is not detrimental to energy efficiency, as discussed in Section~\ref{sec:benchmarking}
The rest of the kernels (CONV, FFT, FIR, and MATMUL) demonstrate a nearly ideal speed-up. 

Vectorization leads to an additional improvement of the speed-up -- between 1.3$\times$ and 2$\times$ -- thanks to the beneficial effects described in Section~1.
Moreover, the improvement derived from vectorization is higher than the parallel speed-up for some applications.
This trend is more evident for FIR, IIR, MATMUL, and KMEANS when passing from 8CL-VECT (8 cores working on vectors) to 16CL (8 cores working on scalars).
This effect is due to the different overheads related to parallelization and vectorization.
As discussed above, IIR and K-MEANS require several synchronization barriers and regions with sequential execution sematic.
Conversely, FIR and MATMUL are amenable to advanced manual vectorization techniques.
For instance, the vector variant of MATMUL reaches a near-ideal improvement vectorizing both input matrices. The efficiency is achieved by unrolling the two inner loops, adding shuffle operations to compute the transpose, and using a dot-product intrinsic to accumulate two products. A similar technique is applied to FIR.
On the other side, the complex multiplication kernel required by FFT requires 7 cycles for scalar data and 10 cycles for vector data; consequently, the maximum gain from vectorization is 1.43$\times$.

\begin{figure}[t]
\begin{center}
\includegraphics[width=0.99\columnwidth]{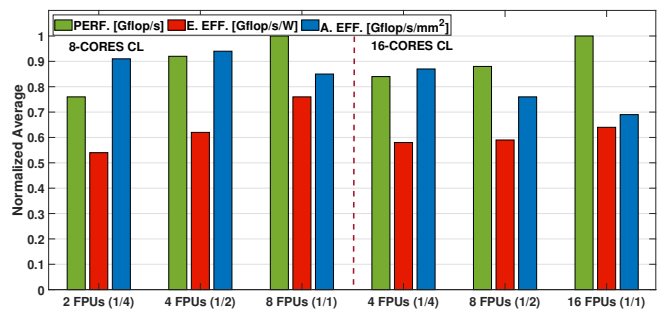} 
%\vspace{-6mm}
\caption{Performance (PERF.), energy efficiency (E. EFF.), and area efficiency (A. EFF.) fixing one pipeline stage and varying the number of FPUs. The values are the average of the normalized results.}
%\vspace{-3mm}
\label{fig:sharing}
\end{center}
\end{figure}
\begin{figure}[t]
\begin{center}
\includegraphics[width=0.99\columnwidth]{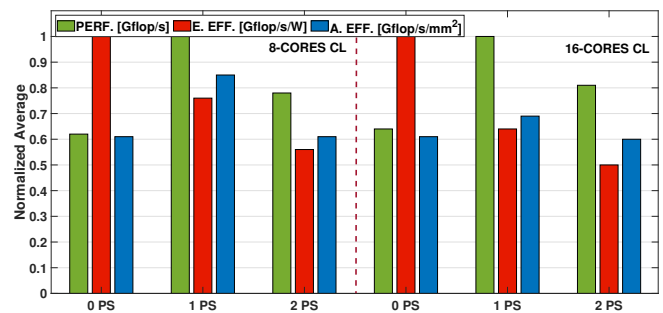} 
%\vspace{-6mm}
\caption{Performance (PERF.), Energy efficiency (E. EFF.), and area efficiency (A. EFF.) fixing a 1/1 sharing factor and varying the pipeline stages (PS). The values are the average of the normalized resultss.}
%\vspace{-3mm}
\label{fig:pipelining}
\end{center}
\end{figure}

\subsubsection{Sharing factor}
Fig.~\ref{fig:sharing} reports average values of performance, energy efficiency, and area efficiency varying the sharing factor. The left part of the figure references 8-cores configurations, the right part 16-cores ones. The number of pipeline stages has been set to one for all experiments, while the number of FPUs corresponds to sharing factors 1/4, 1/2, and 1/1, respectively.

As a general trend, performance grows when increasing the sharing factor.
This increment is more evident passing from 1/4 to 1/2 in 8-cores configurations, and passing from 1/2 to 1/1 in the 16-cores configurations.
%\textbf{TODO DAVIDE How can we explain this effect? It's not so evident, however we should say something.}

The energy efficiency increases with the sharing factor.
This effect is less evident for 16-cores configurations because the contribution of FPUs to the total energy consumption is proportionally lower.
Conversely, the area efficiency increases by reducing the sharing factor from 1/1 to 1/4.
This trend is inverted in the transition from 1/4 to 1/2 with 8 cores. This effect is related to the FP intensity of benchmarks, which is always less than one (as expected in real applications). A 1/2 sharing factor can sustain an FP intensity up to 0.5 with no additional stalls. This value is enough for the requirements of most applications, considering that 0.31 is the average FP intensity of the benchmarks in Table \ref{tab:tab_app}.
On the 16-cores configuration, the number of FPUs to reach the same sharing factor is higher, implying a significant increase of the area; in this case, the best area efficiency corresponds to the minimum sharing factor (1/4).
%
% Please add the following required packages to your document preamble:
% \usepackage{graphicx}
%\begin{tabular}[c]{@{}c@{}}2.10\\ 1.80\\ 0.97\end{tabular}
\begin{table*}[t]
  \centering
  \caption{Comparison with state-of-the-art architectures in high-performance and low-power embedded domain.}
  \label{tab:soa_comparison}
  \setlength{\tabcolsep}{1.5pt}
  \resizebox{\textwidth}{!}{%
    \begin{threeparttable}
      \begin{tabular}{|c|c|c|c|c|c|c|c|c|c|}
      \hline
      & \textbf{Ara} \cite{cavalcante2019ara}
      & \textbf{Hwacha} \cite{lee201445nm}
      & \textbf{Snitch} \cite{zaruba2020snitch}
      & \textbf{Ariane} \cite{zaruba2019floating}
      & \textbf{NTX} \cite{zaruba2019floating}
      & \textbf{Xavier}\tnote{*}
      & \textbf{STM32H7}\tnote{\dag}
      & \textbf{Mr.Wolf} \cite{pullini2019mr}
      & \begin{tabular}[c]{@{}c@{}}\textbf{This work}\\ Best perf. (16c16f1p)\\ Best en. eff. (16c16f0p)\\ Best area eff. (8c4f1p)\end{tabular} \\ \hline
      \textbf{Domain} & High-perf. & High-perf. & High-perf. & High-perf. & High-perf. & Embedded & Embedded & Embedded & Embedded \\ \hline
      \textbf{Technology} & GF 22FDX & 45nm SOI & GF 22FDX & GF 22FDX & GF 22FDX & TSMC 12FFN & 40nm CMOS & 40nm CMOS & GF 22FDX \\ \hline
      \textbf{Voltage (V)} & 0.80\tnote{2} & 0.80\tnote{1} & 0.80\tnote{2} & 0.80\tnote{1} & 0.80\tnote{1} & 0.75\tnote{1} & 1.80 / 1.80\tnote{1} & 1.10\tnote{1} & 0.80 / 0.65 / 0.80\tnote{3}\\ \hline
      \textbf{\begin{tabular}[c]{@{}c@{}}Frequency\\ (GHz)\end{tabular}} & 1.04 & 0.55 & 1.06 & 0.92 & 1.55 & 1.38 & 0.20 / 0.48 & 0.45 & 0.37 / 0.30 / 0.43 \\ \hline
      \textbf{\begin{tabular}[c]{@{}c@{}}Area\\ (mm$^2$)\end{tabular}} & 2.14 & 3.00 & 0.89 & 0.39 & 0.56 & 11.03 & -- & 10.00 & 2.10 / 1.80 / 0.97 \\ \hline
      \textbf{\begin{tabular}[c]{@{}c@{}}Performance\\ (Gflop/s)\end{tabular}} & 64.80 & 3.44 & 14.38 & 2.04 & 18.27 & 153.00 &0.03 / 0.07  & 1.00 & \textbf{2.86} / 2.30 / 1.74 \\ \hline
      \textbf{\begin{tabular}[c]{@{}c@{}}Energy eff.\\ (Gflop/s/W)\end{tabular}} & 81.60 & 25.00 & 103.84 & 33.02 & 110.05 & 52.39 &0.44 / 0.33  & 4.50 & 26.00 / \textbf{81.00} / 23.40 \\ \hline
      \textbf{\begin{tabular}[c]{@{}c@{}}Area eff.\\ (Gflop/s/mm$^2$)\end{tabular}} & 30.34 & 1.14 & 25.83 & 5.23 & 32.63 & 13.84 & -- & 1.70 & 1.50 / 0.60 / \textbf{1.78} \\ \hline
      \textbf{FP formats} & \begin{tabular}[c]{@{}c@{}}float\\ float16\\ bfloat16\\ minifloat\end{tabular} & \begin{tabular}[c]{@{}c@{}}double\\ float\end{tabular} & \begin{tabular}[c]{@{}c@{}}double\\ float\end{tabular} & \begin{tabular}[c]{@{}c@{}}float\\ float16\\ bfloat16\\ minifloat\end{tabular} & float\tnote{\ddag} & \begin{tabular}[c]{@{}c@{}}float\\ float16\end{tabular} & float & float & \begin{tabular}[c]{@{}c@{}}float\\ float16\\ bfloat16\end{tabular} \\ \hline
      \textbf{\begin{tabular}[c]{@{}c@{}}Programming\\ interface\end{tabular}} & ISA extension & ISA extension & ISA extension & ISA extension & \begin{tabular}[c]{@{}c@{}}Memory-mapped\\ configuration\end{tabular} & Base ISA & Base ISA & Base ISA & ISA extension \\ \hline
      \textbf{\begin{tabular}[c]{@{}c@{}}Execution\\ model\end{tabular}} & \begin{tabular}[c]{@{}c@{}}SIMD vector\\ unit\\ (accelerator)\end{tabular} & \begin{tabular}[c]{@{}c@{}}SIMT\\ vector-thread\\ unit\\ (accelerator)\end{tabular} & \begin{tabular}[c]{@{}c@{}}Loop-buffers for\\ tensor streaming\\ (accelerator)\end{tabular} & \begin{tabular}[c]{@{}c@{}}SIMD\\ processor\end{tabular} & \begin{tabular}[c]{@{}c@{}}Loop-buffers for\\ tensor streaming\\ (accelerator)\end{tabular} & \begin{tabular}[c]{@{}c@{}}SIMT\\ vector-thread\\ unit\\ (accelerator)\end{tabular} & Processor & \begin{tabular}[c]{@{}c@{}}Multi-core\\ processor\end{tabular} & \begin{tabular}[c]{@{}c@{}}Multi-core\\ processor\end{tabular} \\ \hline
      \textbf{\begin{tabular}[c]{@{}c@{}}Compiler\\ support\end{tabular}} & Yes & Yes (OpenCL) & Partial (inline ASM) & Yes & No & Yes (CUDA) & Yes & Yes & Yes \\ \hline
      \end{tabular}%
      \begin{tablenotes}
        \item[*] Numbers extracted from \cite{zaruba2020snitch}.
        \item[\dag] Measurements taken on a NUCLEO\-H743ZI development board executing a 128$\times$128. matrix multiplication.
        \item[1] Silicon measurements.
        \item[2] Post-layout simulation using \emph{typical} frequency.
        \item[3] Post-layout simulation using \emph{worst-case} frequency.
        \item[\ddag] Higher internal accumulation precision with float results.
      \end{tablenotes}
    \end{threeparttable}%
  }
\end{table*}

\subsubsection{Pipelining}
Fig.~\ref{fig:pipelining} shows average values of performance, energy efficiency, and area efficiency varying the number of pipeline stages.
The support for pipelining improves performance since this technique allows for increasing the operating frequency of the transprecision cluster.
Conversely, performance degrades with two pipeline stages. Even if we can increase the operating frequency, we observe an increment in the number of cycles due to the contentions on the write ports of the register file.
Configuring the FPU for two pipeline stages, a write-back stall may happen when a load/store post-increment operation or an integer operation arrives right after an FP operation. 
For instance, when we have the valid signal for the FP operation in the first cycle, and then a load/store post-increment request in the second clock cycle before storing the FP results, the FPU must wait until the other instructions end, resulting in a stall for the use of the port. 
There are no contentions with no pipeline stages because there is a dedicated port for the FPU.

In all cases, energy efficiency decreases by incrementing the number of pipeline stages since the design makes the logic more complicated.
Finally, area efficiency follows a trend very similar to performance. The area required to enable one-stage pipelining leads to a considerable benefit while adding additional area for a second stage is not so convenient.
This trend is less evident on 16-cores configurations since the impact of pipeline logic is less significant.
 \section{Comparison with the SoA}
\label{sec:soa}
Table~\ref{tab:soa_comparison} depicts a comparison with SoA platforms with FP support in high-performance and embedded domains.
The number of FP operations has been measured by executing a single-precision matrix multiplication on all the platforms. We have considered three configurations of the transprecision cluster (reported in the TP column), corresponding to the best performance, the best energy efficiency, and the best area efficiency. In the domain of low-power embedded systems, our solution outperforms a single-core Cortex core and the Mr.Wolf multi-core cluster in all metrics. The Tegra Xavier SoC contains eight streaming multiprocessors (SMs) composed of four execution units. Each execution unit includes 16 single-precision FPUs sharing a register file and an instruction cache. The transprecision cluster is 53\% better than an SM in terms of energy efficiency. As regards performance, a single execution unit is 13$\times$ faster.

As expected, the absolute performance and area efficiency of platforms in the high-performance domain is higher than the transprecision cluster due to the higher operating frequencies. In our design, we consider the worst-case corner for the computation of the operating frequency, while the other solutions report silicon results or typical corners, which is somehow penalizing for us. Nevertheless, our solution outperforms an Ariane and is comparable with a Hwacha vector processor. The energy efficiency of the transprecision cluster is comparable with Snitch, NTX, and Ara, despite these architectures are heavily specialized for FP intensive computations. This outcome is due to three main factors. First, operating at low voltage in near-threshold operation makes the transprecision cluster very power efficient. Second, the best solution is not to adopt pipelining, so it does not pay the energetic overhead of pipeline logic. Third, the support to FMA operations increments by 2$\times$ the number of operations performed per cycle and is highly beneficial.

Finally, compared to most energy-efficient solutions in Table~\ref{tab:soa_comparison}, the proposed cluster provides full compiler support and flexibility typical of high-level parallel programming models such as OpenMP, not requiring programmers to use low-level accelerator-centric interfaces such as OpenCL or memory-mapped APIs, or even lower-level abstractions (e.g., inline assembly).
This support is a key requirement for the wide adoption of these solutions for near-sensor computing.
 \section{Conclusion}
\label{sec:conclusion}
In this paper, we have described the design of a transprecision cluster for near-sensors computing, providing a full specification of its main components and a software ecosystem to execute real applications.
We have performed a design space exploration on an FPGA emulator varying the number of cores, the number of FPUs, and the pipeline stages.
A set of experiments on near-sensor applications and an analysis on post P\&R models have allowed us to identify the most efficient configurations.

Our experimental results show that configurations with 16 cores and private FPUs are the best solution in terms of performance and energy efficiency, while a cluster with 8 cores and 4 shared FPUs remains the best solution for area efficiency.
Moreover, these results highlight two important outcomes.
First, energy efficiency is not affected by the effectiveness of parallelization techniques since the platform provides effective power-saving policies to turn off cores that are not active.
Second, the trend for energy efficiency is different from area efficiency; in the design space that we are considering, these metrics are related but do not scale with a fixed rate.
To conclude, one pipeline stage is the solution providing the best compromise in most configurations; no pipelining can be beneficial only when energy saving is a high-priority constraint.
These outcomes provide a useful insight to system designers and engineers, and the guidelines derived by this exploration can steer the design of future near-sensor computing platforms in a wide range of application domains.

 \bibliographystyle{IEEEtran}

 \bibliography{CONTENTS/bibliography}

% For peer review papers, you can put extra information on the cover
% page as needed:
% \ifCLASSOPTIONpeerreview
% \begin{center} \bfseries EDICS Category: 3-BBND \end{center}
% \fi
%
% For peerreview papers, this IEEEtran command inserts a page break and
% creates the second title. It will be ignored for other modes.
\IEEEpeerreviewmaketitle

% if have a single appendix:
%\appendix[Proof of the Zonklar Equations]
% or
%\appendix  % for no appendix heading
% do not use \section anymore after \appendix, only \section*
% is possibly needed

% use appendices with more than one appendix
% then use \section to start each appendix
% you must declare a \section before using any
% \subsection or using \label (\appendices by itself
% starts a section numbered zero.)
%

%\appendices
%\section{Proof of the First Zonklar Equation}
%Appendix one text goes here.

% you can choose not to have a title for an appendix
% if you want by leaving the argument blank
%\section{}
%Appendix two text goes here.

% use section* for acknowledgment
% \ifCLASSOPTIONcompsoc
%   % The Computer Society usually uses the plural form
% \section*{Acknowledgments}
% \else
%   % regular IEEE prefers the singular form
%   \section*{Acknowledgment}
% \fi

% Can use something like this to put references on a page
% by themselves when using endfloat and the captionsoff option.
\ifCLASSOPTIONcaptionsoff
  \newpage
\fi

\vspace{-30pt}
\begin{IEEEbiography}[{\includegraphics[width=1in,height=1.25in,clip,keepaspectratio]{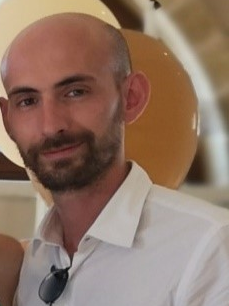}}]{Fabio Montagna}
received the Ph.D. degree in Electrical, Electronic and Information Engineering from the University of Bologna, Bologna, Italy, in 2020. He is currently working as Research Fellow at DISI, University of Bologna, Bologna, Italy. His main research topic is energy-efficient parallel architectures for ultra-low power biosignal processing. His research interests include embedded wearable and implantable systems, parallel computing, signal processing, and machine learning.
\end{IEEEbiography}
\vspace{-30pt}
\begin{IEEEbiography}[{\includegraphics[width=1in,height=1.25in,clip,keepaspectratio]{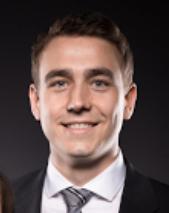}}]{Stefan Mach}
received his B.Sc. and M.Sc. degree from the Swiss Federal Institute of Technology Zurich (ETHZ), Switzerland, where he is currently pursuing a Ph.D. degree.
Since 2017, he has been a research assistant with the Integrated Systems Laboratory at ETHZ.
His research interests include transprecision computing, computer arithmetics, and energy-efficient processor architectures.
\end{IEEEbiography}
\vspace{-30pt}
\begin{IEEEbiography}[{\includegraphics[width=1in,height=1.25in,clip,keepaspectratio]{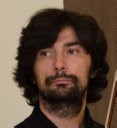}}]{Simone Benatti} received the Ph.D. degree in Electrical Engineering and Computer Science from the University of Bologna, in 2016. He currently holds a postdoctoral position with the University of Bologna. His research interests include energy efficient embedded wearable systems for advanced Human Computer Interaction and algorithms for edge computing. In this field, he has authored/coauthored more than 40 papers in international peer-reviewed conferences and journals. Previously, he worked 8 years as an Electronic Designer and R\&D Engineer of electromedical devices.
\end{IEEEbiography}
\vspace{-30pt}
\begin{IEEEbiography}[{\includegraphics[width=1in,height=1.25in,clip,keepaspectratio]{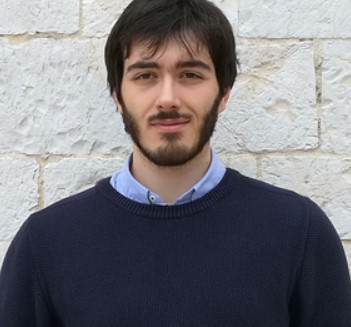}}]{Angelo Garofalo}
received the B.Sc and M.Sc. degree in electronic engineering from the University of Bologna, Bologna, Italy, in 2016 and 2018 respectively. He is currently working toward his Ph.D. degree at DEI, University of Bologna, Bologna, Italy. His main research topic is Hardware-Software design of ultra-low power multiprocessor systems on chip. His research interests include Quantized Neural Networks, Hardware efficient Machine Learning, transprecision computing, and energy-efficient fully-programmable embedded architectures.
\end{IEEEbiography}
\vspace{-30pt}
\begin{IEEEbiography}[{\includegraphics[width=1in,height=1.25in,clip,keepaspectratio]{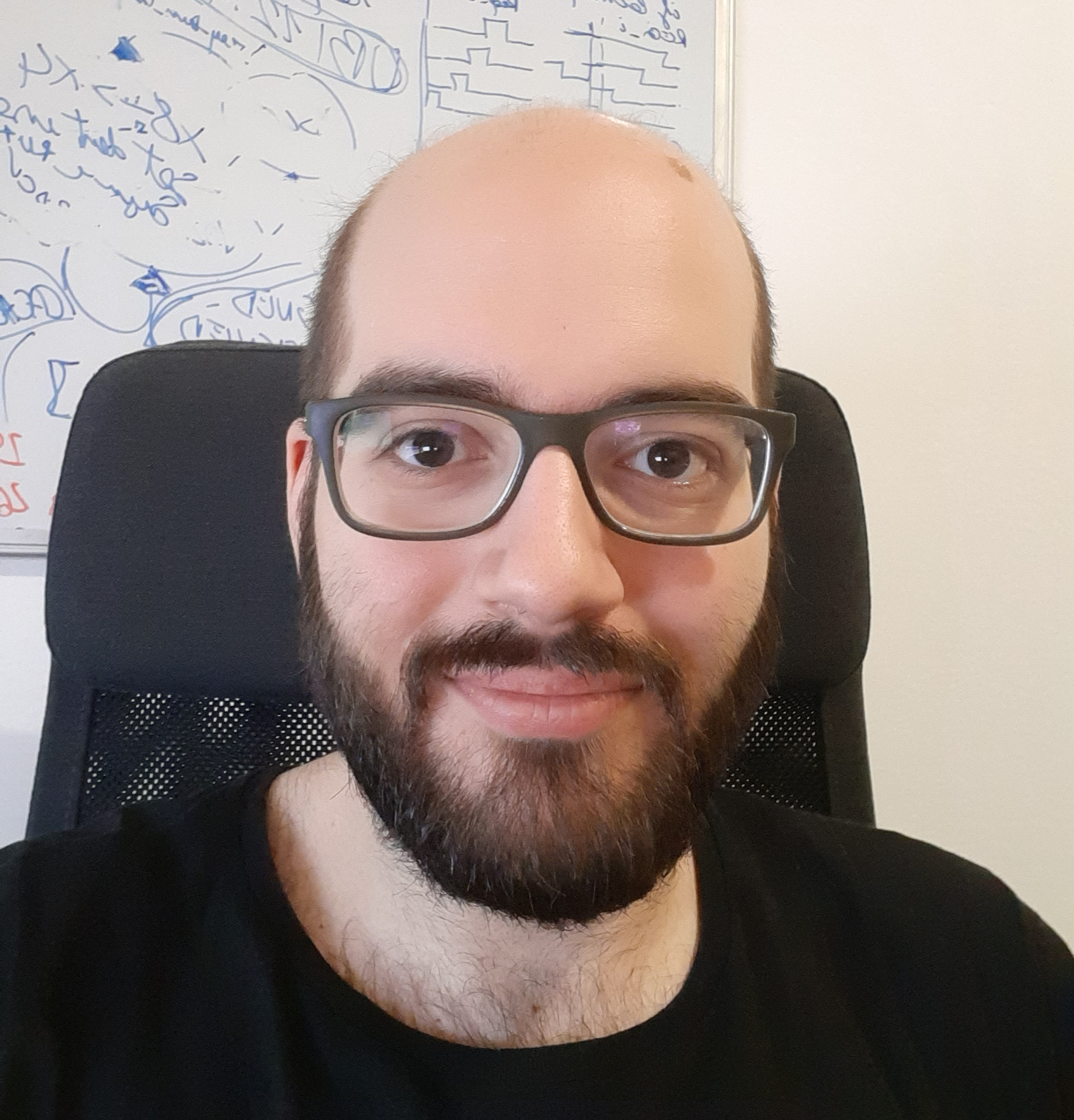}}]{Gianmarco Ottavi}
is a researcher at University of Bologna (Italy) in the department of Electrical, Electronic and Information Engineering (DEI). His current resarch topics are on developing Ultra Low Power embedded systems based on RISC-V Instruction Set Architecture.
\end{IEEEbiography}
\vspace{-30pt}
\begin{IEEEbiography}[{\includegraphics[width=1in,height=1.25in,clip,keepaspectratio]{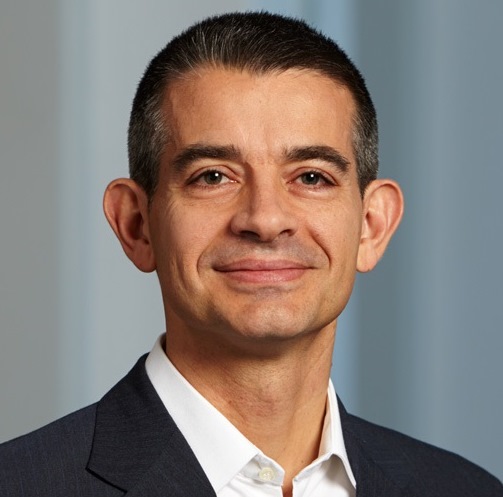}}]{Luca Benini}
holds the chair of digital Circuits and systems at ETHZ and is Full Professor at the Universita di Bologna.
Dr. Benini’s research interests are in energy-efficient computing systems design, from embedded to high-performance.
He has published more than 1000 peer-reviewed papers and five books.
He is a Fellow of the ACM and a member of the Academia Europaea.
He is the recipient of the 2016 IEEE CAS Mac Van Valkenburg Award and the 2020 EDAA Achievement Award.
\end{IEEEbiography}
\vspace{-30pt}
\begin{IEEEbiography}[{\includegraphics[width=1in,height=1.25in,clip,keepaspectratio]{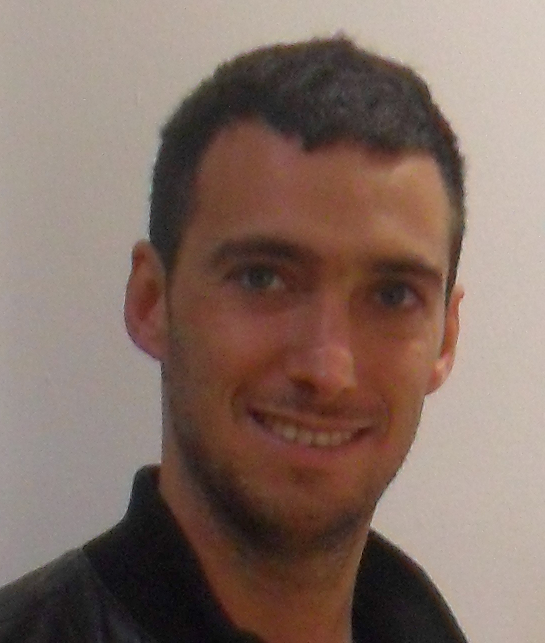}}]{Davide Rossi}
received the Ph.D. degree from the University of Bologna, Bologna, Italy, in 2012.
He has been a Post-Doctoral Researcher with the Department of Electrical, Electronic and Information Engineering “Guglielmo Marconi,” University of Bologna, since 2015, where he is currently an Assistant Professor. His research interests focus on energy-efficient digital architectures. In this field, he has published more than 100 papers in international peer-reviewed conferences and journals. He is recipient of Donald O. Pederson Best Paper Award 2018, 2020 IEEE TCAS Darlington Best Paper Award, 2020 IEEE TVLSI Prize Paper Award.
\end{IEEEbiography}
\vspace{-30pt}
\begin{IEEEbiography}[{\includegraphics[width=1in,height=1.25in,clip,keepaspectratio]{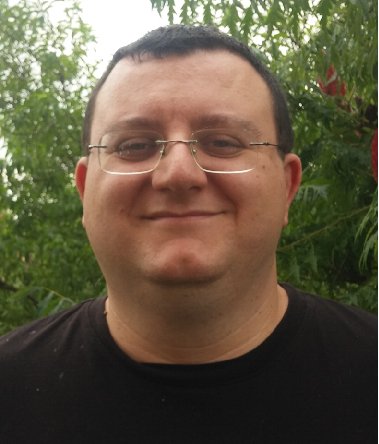}}]{Giuseppe Tagliavini}
received the Ph.D. degree in electronic engineering from the University of Bologna, Bologna, Italy, in 2017.
He is currently an Assistant Professor with the Department of Computer Science and Engineering (DISI) at the University of Bologna. He has co-authored over 30 papers in international conferences and journals. His research interests include parallel programming models for embedded systems, run-time optimization for multicore and many-core accelerators, and design of software stacks for emerging computing architectures.
\end{IEEEbiography}
%
%

% You can push biographies down or up by placing
% a \vfill before or after them. The appropriate
% use of \vfill depends on what kind of text is
% on the last page and whether or not the columns
% are being equalized.

%\vfill

% Can be used to pull up biographies so that the bottom of the last one
% is flush with the other column.
%\enlargethispage{-5in}

% that's all folks
\end{document}